\documentclass[fleqn,10pt]{wlscirep}
\usepackage[utf8]{inputenc}
\usepackage[T1]{fontenc}
\usepackage{color}
\usepackage{float}
\title{Full-plane persistent spin textures with cubic order intrinsic and anisotropic band splitting in bulk Lead-free materials}

\author[1*]{Sajjan Sheoran}
\author[1]{Preeti Bhumla}
\author[1*]{Saswata Bhattacharya}
\affil[1]{Department of Physics, Indian Institute of Technology Delhi, New Delhi 110016, India}

\affil[*]{sajjan@physics.iitd.ac.in [SS]}
\affil[*]{saswata@physics.iitd.ac.in [SB]}

\begin{abstract}
	Spin-orbit coupling (SOC) effects occurring in noncentrosymmetric materials are known to be responsible for nontrivial spin configurations and a number of emergent physical phenomena such as electrical control of spin degrees of freedom and spin-to-charge conversion. The materials preserving a uniform spin configuration in the momentum-space, known as persistent spin texture (PST), provide long carrier spin lifetimes through persistent spin helix (PSH) mechanism. However, most of the PST studied till now are attributed to the linear in \textbf{\textit{k}} splitting and cease to exist locally around certain high-symmetry-point of first Brillouin Zone (FBZ). The persistent spin textures with purely cubic spin splittings have drawn attention owing to unique benefits in spin transport. Here, by using the relativistic first-principles calculations supplemented with \textbf{\textit{k.p}} analysis, we report the emergence of purely cubic splitting (PCS) belonging to $D_{3h}$ point group, which is enforced by in-plane mirror and three-fold rotation operations. In addition, the in-plane mirror symmetry operation sustains the PST in larger region (i.e. full planes) of FBZ alongside giant spin splitting. Our results also demonstrate how application of uniaxial strain could be envisaged to tune the magnitude of the PCS, preserving the PST. The observed PSTs provide a route to non-dephasing spin transport with larger spin-Hall conductivity, thus offering a promising platform for future spintronics devices.

\end{abstract}
\begin{document}

\flushbottom
\maketitle
%
%
\thispagestyle{empty}


\section*{Introduction}
In non-magnetic solids, one can naively expect the electron energy bands having up and down spins to be degenerate in absence of magnetic field. However, in systems that break spatial inversion symmetry, spin-orbit coupling (SOC) generates a momentum \textbf{\textit{k}}-dependent spin-orbit field (SOF), $\Omega(\textit{\textbf{k}})\propto \textit{\textbf{E}}\times \textit{\textbf{k}}$, where $\textit{\textbf{E}}$ is the electric field induced by the inversion asymmetry of the crystal\cite{dresselhaus1955spin, bychkov1984properties, moriya2014cubic, nakamura2012experimental}. The SOF lifts the spin band degeneracy, resulting into non-trivial \textbf{\textit{k}}-dependent spin textures mainly through Rashba and Dresselhaus effects~\cite{gmitra2016first, marchenko2012giant,stranks2018influence, bihlmayer2015focus,tao2017reversible,di2013electric}. The charges with opposite spin have opposite directions in momentum space, thus they show spin-selective transport. In particular, linear Rashba (LR) effect has attracted much attention since it can manipulate the spin degrees of freedom electrically and produce non-equilibrium spin polarization~\cite{picozzi2014ferroelectric, da2016rashba, varignon2019electrically}. These phenomena are steering a new way in various intriguing fields, including spintronics and topological matter~\cite{ moriya2014cubic, nakamura2012experimental,gmitra2016first, marchenko2012giant,stranks2018influence, bihlmayer2015focus,tao2017reversible,di2013electric, tao2018persistent, djani2019rationalizing, picozzi2014ferroelectric, varignon2019electrically, plekhanov2014engineering, arras2019rashba,autieri2019persistent,stroppa2014tunable,ishizaka2011giant, bandyopadhyay2020origin, koralek2009emergence, manchon2015new , jia2020persistent }. 

\begin{figure}[h]
		\begin{center}
		\includegraphics[width=10cm]{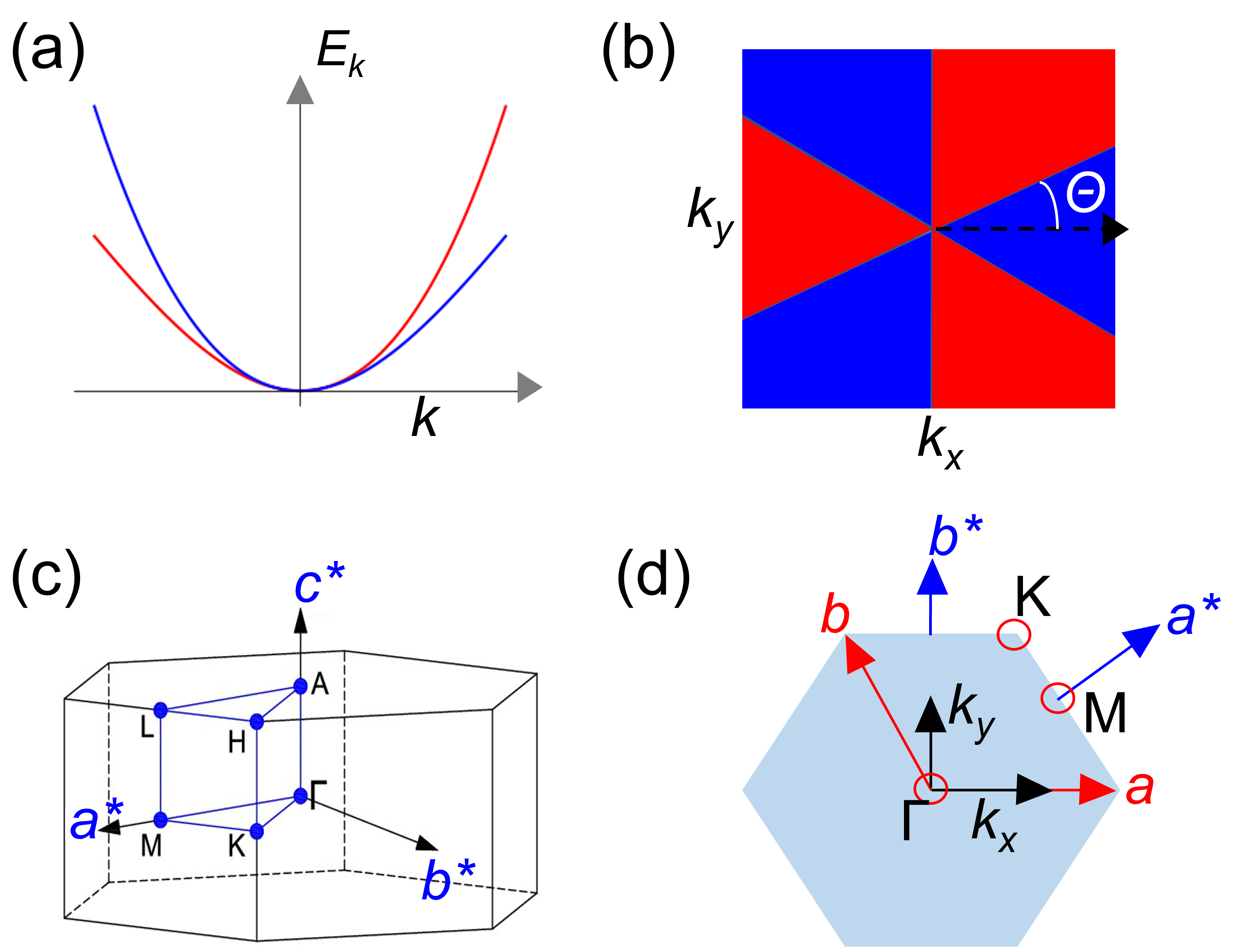}
		\caption{(a) Schematic illustration of band structure due to purely cubic spin splitting. Here, the band energies $E_\textbf{\textit{k}}$ vary with \textbf{\textit{k}} in the form, $E_\textbf{\textit{k}} = \alpha \textbf{\textit{k}}^2 \pm  \eta \textbf{\textit{k}}^3$. (b) Schematic persistent spin texture due to cubic spin splittings. The color represents the out-of-plane (\textit{z}-) component of spin textures (note that in-plane ($x$-, $y$-) spin components is zero). Here, the red and blue colors denote the spins in the state $\lvert \uparrow \rangle $ and $\lvert \downarrow \rangle $, respectively. Here, $\lvert \uparrow \rangle $ and $\lvert \downarrow \rangle $ are the eigenstates of $\sigma_z$ with eigenvalues $+1$ and $-1$, respectively. $\Theta$ is the angular difference between line separating the sectors with eigenvalues $+1$ and $-1$, and $k_x=0$. (c) First Brillouin zone (BZ) of the hexagonal structure. Blue lines highlight the path used for band structure calculation. (d) BZ of the hexagonal structure within the ($k_x - k_y$) plane, where $k_x$ and $k_y$ are the cartesian reciprocal lattice vectors. The lattice vectors are denoted by $a$ and $b$, whereas the reciprocal ones
			are represented by $a$* and $b$*, respectively.}
		\label{fig1}
	\end{center}
	\end{figure}

In general, these spin splittings are well explained by the linear Rashba (LR), linear Dresselhaus (LD) or some unique combination of both LR and LD~\cite{ bihlmayer2015focus,tao2017reversible,di2013electric, tao2018persistent, djani2019rationalizing, picozzi2014ferroelectric, da2016rashba, varignon2019electrically, arras2019rashba,stroppa2014tunable,ishizaka2011giant, bandyopadhyay2020origin, koralek2009emergence, manchon2015new , jia2020persistent}. In linear spin splitting including LR or LD, energy levels are dispersed with crystal momentum (\textbf{\textit{k}}) having the relation $E_{\textbf{\textit{k}}} = \alpha \textbf{\textit{k}}^2 \pm  \tau \textbf{\textit{k}}$, where $\alpha$ and $\tau$ are the  effective mass and linear splitting terms, respectively. The LR and LD can be described by SOC perturbed Hamiltonians given by, $H_{LR}=\alpha_R(\sigma_x k_y-\sigma_y k_x)$ and $H_{LD}=\alpha_D(\sigma_x k_y-\sigma_y k_x)$, respectively~\cite{tao2017reversible}. Here, $\alpha_R$ and $\alpha_D$ are the constants signifying the strength of LR  and LD effects, respectively, and \textbf{$\sigma$} is the Pauli matrices vector. However, the SOF arising from LR or LD will introduce two kinds of spin dephasing mechanisms viz., Elliott-Yafet spin relaxation and Dyakonov-Perel spin dephasing, which limit the experimental realization of spin currents. The PSTs where spin configurations become independent of the crystal momentum (\textbf{\textit{k}}), enable a route to overcome spin dephasing and provide non dissipative spin transport~\cite{tao2018persistent,djani2019rationalizing,autieri2019persistent,jia2020persistent,schliemann2003nonballistic,schliemann2017colloquium}. In the regime of linear splitting, PST can be obtained by tuning the strength of LR and LD effects, such that they compensate each other ($\alpha_R = \alpha_D$). In addition, PST with linear spin splitting can also be enforced by nonsymmorphic space-group symmetry of the crystal and therefore, not easily broken~\cite{tao2018persistent}. In recent times, cubic Rashba (CR) and cubic Dresselhaus (CD) effects have started to gain scrutiny due to unique benefits to spin transport like larger spin-Hall conductivity as compared to linear spin splitting effects and  also provide spin transport in a system with higher spin~\cite{moriya2014cubic,nakamura2012experimental,gmitra2016first,shanavas2016theoretical,lin2019interface, marinescu2017cubic, schliemann2005spin}. The purely cubic effects (CR or CD) can be distinguished from linear effects (LR or LD) by dispersion curve $E_\textbf{\textit{k}} = \alpha \textbf{\textit{k}}^2 \pm  \eta \textbf{\textit{k}}^3$ (see Figure~\ref{fig1}a). The CR and CD effects are described by the SOC Hamiltonians given by, $H_{CR}=(k_x-ik_y)^3(\sigma_x+i\sigma_y)-(k_x+ik_y)^3(\sigma_x-i\sigma_y)$ and $H_{CD}=(k_xk_z^2-k_xk_y^2)\sigma_x + (k_yk_x^2-k_yk_z^2)\sigma_y + (k_zk_y^2-k_zk_x^2)\sigma_z$, respectively. The crucial experimental discoveries related to CR contain SrTiO$_3$(001) surface~\cite{nakamura2012experimental}, asymmetric oxide heterostructure LaAlO$_3$/SrTiO$_3$/LaAlO$_3$~\cite{lin2019interface}, strained-Ge/SiGe quantum well~\cite{moriya2014cubic}, surfaces of antiferromagnets GdIr$_2$Si$_2$ and TbRh$_2$Si$_2$~\cite{schulz2021classical,usachov2020cubic}. The CR effect is mainly reported in surfaces, heterostructures and interfaces.

The PST in the regime of PCS (without mixing with the splitting linear in \textbf{\textit{k}}) in bulk materials are of great importance due to aforementioned unique benefits. Recently, symmetry enforced PCS was predicted for bulk materials by deriving the two-band \textit{\textbf{k.p}} model of all 21 polar point group symmetries, having $C_{3h}$, $D_{3h}$ and $T_d$ ($\overline{6}$, $\overline{6}m2$ and $\overline{4}3m$  in Hermann-Mauguin notation) as the associated point group symmetries~\cite{zhao2020purely}. However, spilitting in $T_d$ point group corresponds to trivial CD, which can be seen in zinc-blende type crystal structures~\cite{wang2007dresselhaus}. In addition, it was also shown that the crystals with $C_{3h}$ and $D_{3h}$ point group symmetries show PST with PCS around the $\Gamma$ point (see Figure~\ref{fig1}b). These types of PST are preserved by the symmetry and thus, robust to spin-independent impurities and not easily destroyed. These findings are reported for Ge$_3$Pb$_5$O$_{11}$, Pb$_7$Br$_2$F$_{12}$ and Pb$_7$Cl$_2$F$_{12}$ having $P\overline{6}$ space group symmetry (with $C_{3h}$ associated point group symmetry)~\cite{zhao2020purely}. However, Ge$_3$Pb$_5$O$_{11}$ is only stable at temperatures above 450 K and conduction band minima (CBm) of Pb$_7$Br$_2$F$_{12}$ and Pb$_7$Cl$_2$F$_{12}$ do not lie in the vicinity of $\Gamma$ point~\cite{iwata1973neutron,zhao2020purely}. Alongside these limitations, presence of toxic element Pb is also a concern. Therefore, finding the novel structures which can circumvent these limitations and supporting PST with PCS are of great importance.

In the present work, using the first-principles density-functional theory (DFT) calculations, which are additionally supplemented by effective model Hamiltonian derived from the \textbf{\textit{k.p}} invariant method, we reveal the emergence of PCS in large family of materials with $D_{3h}$ point group symmetry. We find that the in-plane mirror and three-fold rotation symmetry of the crystal lead to the PCS around $\Gamma$ and A points in the Brillouin zone. In addition, the PST are observed in a larger region and preserved in whole $\Gamma-$M$-$K  plane (see Figure~\ref{fig1}(c)-(d) for the plane).  More importantly,  K$_3$Ta$_3$B$_2$O$_{12}$ exhibits giant spin splitting, which is particularly observed around the $\Gamma$, M, and A points in the proximity of the CBm. Our calculations also show that magnitude of the spin splitting can be tuned with the application of uniaxial strain preserving the PST.
\section*{Results}
	\begin{figure}[h]
	\begin{center}
		\includegraphics[width=15cm]{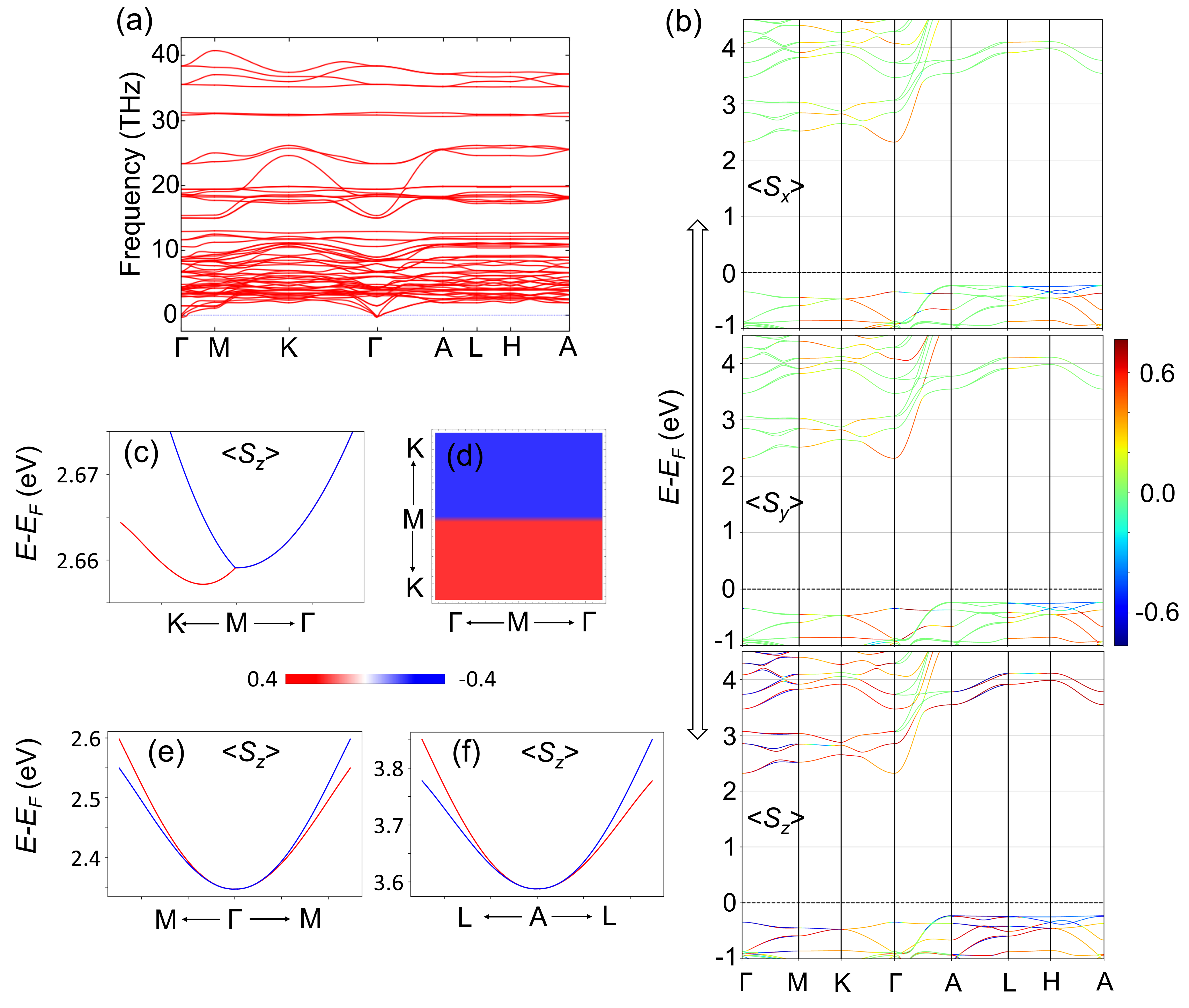}
		\caption{(a) Calculated phonon dispersion for K$_3$Ta$_3$B$_2$O$_{12}$. (b) Spin-resolved band structures along high symmetry path are shown. The color bars represent the expectation values of the $S_x$, $S_y$ and $S_z$ spin components.  (c) The calculated spin resolved band structure of lower conduction bands along the path K$-$M$-$$\Gamma$. (d) Spin texture of lower conduction band around the M point in $k_z=0$)plane. Note that there should be two spin-textures associated with two bands. Only one branch associated with the outer band is shown here; the inner branch has opposite spin orientation with respect to outer branch at every $(k_x,k_y)$ point. The spin resolved band structures around (e) $\Gamma$ point (along  M$-$$\Gamma-$M) and (f) A point (along  L$-$A$-$L).}
		\label{fig_band}
	\end{center}
\end{figure}
\subsection*{Symmetry analysis and electronic band structures}
We have screened lead-free experimentally synthesized non-magnetic materials having band gap greater than 1.0 eV with  space groups $P\overline{6}$, $P\overline{6}m2$, $P\overline{6}2m$, $P\overline{6}2c$ and $P\overline{6}c2$, from AFLOW~\cite{curtarolo2012aflow} and Materials Project~\cite{jain2013commentary} data repositories. There is rather a large family of materials which crystallize in these space groups, and can show PST with PCS (see section I of supplementary information (SI)). In particular, we are interested in materials, which are dynamically stable, having suitable band edge positions, wide band gap and exhibit giant splitting. Thus, we have particularly focused on K$_3$Ta$_3$B$_2$O$_{12}$ ($P\overline{6}m2$), KTaGe$_3$O$_{9}$ ($P\overline{6}2c$) with $D_{3h}$ point group symmetry and Sr$_7$Br$_2$H$_{12}$ ($P\overline{6}$) with $C_{3h}$ point group symmetry. The relaxed crystal structures and atomic positions are shown in section II of SI. Our relaxed crystal structures are in good agreement with the experimentally synthesized crystal structures. The calculated lattice parameters ($a$, $c$) deviate from experimental ones~\cite{abrahams1981piezoelectric, choisnet1972nouveaux,reckeweg2008syntheses}  by (1.2\%, 1.2\%), (0.3\%, 2.1\%) and (1.5\%, 1.4\%) for K$_3$Ta$_3$B$_2$O$_{12}$, KTaGe$_3$O$_{9}$ and Sr$_7$Br$_2$H$_{12}$, respectively. The dynamical stability of these materials are tested by analyzing the phonon spectra. Figure~\ref{fig_band}a shows the phonon spectra for  K$_3$Ta$_3$B$_2$O$_{12}$ along the high symmetry path given in Figure~\ref{fig1}c (see section III of SI for phonon plots of KTaGe$_3$O$_{9}$ and Sr$_7$Br$_2$H$_{12}$). The absence of negative frequencies in the phonon dispersion plots confirms the dynamical stability of these structures at low temperature.

Firstly, we have calculated the band structures using PBEsol exchange correlation( $\epsilon_{xc}$)  functional with inclusion of SOC. Figure~\ref{fig_band}bs show spin projected band structure of K$_3$Ta$_3$B$_2$O$_{12}$. A band gap of 2.51 eV is estimated with CBm and valence band maximum (VBM) at $\Gamma$ and A points, respectively. Since PBEsol is known to underestimate the band gap, it was also estimated using more sophisticated hybrid $\epsilon_{xc}$ functional HSE06. The band gap has increased to 4.01 eV without altering the band edge positions. The band gaps are found to be 3.42 eV  and 3.51 eV for KTaGe$_3$O$_{9}$ and Sr$_7$Br$_2$H$_{12}$, respectively (see section IV in the SI). We have found that CBm of  KTaGe$_3$O$_{9}$ is located at $\Gamma$ point, whereas CBm of Sr$_7$Br$_2$H$_{12}$ lies at H point. The VBM is located at  M point and $\Gamma$ point for  KTaGe$_3$O$_{9}$ and Sr$_7$Br$_2$H$_{12}$, respectively.  The HSE06 band gaps come out to  be 4.74 eV and 4.84 eV for KTaGe$_3$O$_{9}$ and Sr$_7$Br$_2$H$_{12}$, respectively.
\begin{table}[h]
	\begin{center}
		\caption {The transformations of ($\sigma_x$, $\sigma_y$, $\sigma_z$) and ($k_x$, $k_y$, $k_z$) with respect to the generators of the $C_s$, $D_{3h}$ and $C_{3h}$ point group and time-reversal operator ($T$). First row shows the point group operations and the corresponding point group symmetries. Note that the generators are enough to form the whole group. Hence, only these generators along with time-reversal $T=i\sigma_yK$ operation ($K$ is complex conjugation operator) are considered to construct the \textbf{\textit{k.p}} model. The last row shows the terms which are invariant under point group operation. Note that we have included the terms upto cubic in \textbf{\textit{k}} and higher order contributions are found to be insignificant.}
		\begin{tabular}{p{2.0cm}p{3.5cm}p{4.0cm}p{4.0cm}p{2.0cm}}
			\hline
			\hline
			Operations & $C_{3z}=e^{{-i\pi}/{3\sigma_z}}$ & $M_{xz}=i\sigma_y$ & $M_{xy}=i\sigma_z$& $T=i\sigma_yK$           \\ 
			
			& ($D_{3h}$, $C_{3h}$) & ($D_{3h}$, $C_s$)  & ($D_{3h}$, $C_{3h}$, $C_s$)&           \\ \hline \hline
			$\;\;\;\,k_x$       &  $-k_x/2+\sqrt{3}k_y/2$  &$\;\;\;\, k_x$ & $\;\;\;\,k_x$& $-k_x$        \\ 
			$\;\;\;\,k_y$       &  $-\sqrt{3}k_x/2-k_y/2$  & $-k_y$ &$\;\;\;\,k_y$& $-k_y$      \\ 
			$\;\;\;\,k_z$       &  $\;\;\;\,k_z$  & $\;\;\;\,k_z$ &$-k_z$& $-k_z$       \\ 
			$\;\;\;\,\sigma_x$  &  $-\sigma_x/2+\sqrt{3}\sigma_y/2$  & $-\sigma_x$   &$-\sigma_x$& $-\sigma_x$   \\ 
			$\;\;\;\,\sigma_y$  &  $-\sqrt{3}\sigma_x/2-\sigma_y/2$  & $\;\;\;\,\sigma_y$  &$-\sigma_y$ & $-\sigma_y$   \\ 
			$\;\;\;\,\sigma_z$  &  $\;\;\;\,\sigma_z$  & $-\sigma_z$   &$\;\;\;\,\sigma_z$& $-\sigma_z$  \\ \hline\hline
			&  $(k_x\sigma_y-k_y\sigma_x$),  & $k^m_ik_y\sigma_x$, $k^m_ik_y\sigma_z$,    & $k^m_ik_x\sigma_z$, $k^m_ik_y\sigma_z$,& $k_i\sigma_j $ \\
			Invariants            &  $k_y(3k_x^2-k_y^2)\sigma_z$,  & $k^m_ik_x\sigma_y$, $k^m_ik_z\sigma_y$   &$k^m_ik_z\sigma_x$, $k^m_ik_z\sigma_y$ & ($i,j=x,y,z$) \\ 
			& $k_x(k_x^2-3k_y^2)\sigma_z$ & ($i=x,y,z$; $m=0,2$)  & ($i=x,y,z$; $m=0,2$)& \\ \hline \hline
		\end{tabular}
		\label{tran}
	\end{center}
\end{table}
\subsection*{ Spin splitting and effective \textit{k.p} Hamiltonian}
  Inclusion of SOC leads to splitting of  bands throughout the Brillouin zone. In K$_3$Ta$_3$B$_2$O$_{12}$, splitting is predominantly present for lower conduction band along $\Gamma$-M, A-L directions (highlighted by black circles in Figure ~\ref{fig_band}b) and nearly absent for upper valence band. This is due to the fact that lower conduction band mainly consists of heavier Ta-5d orbitals, whereas O-2p orbitals contribute to upper valence band. Along the path $\Gamma$-M-K-$\Gamma$ ($k_z$=0 plane), it is clearly seen that the band dispersion is fully characterized by the out-of-plane spin component. In contrary, the in-plane spin component is almost zero along that path. The PST is observed in full plane and different from widely reported PST, which occurs locally around high-symmetry-point or along lines of FBZ. The in-plane mirror symmetry of crystal ($M_{xy}$) enforces spin components to hold the relation $(S_x,S_y,S_z)\rightarrow (-S_x,-S_y,S_z)$, which satisfies only when in-plane components are zero. This leads to unidirectional SOF, which is out-of-plane, considered to be the $z$-axis. Figure ~\ref{fig_band}c and ~\ref{fig_band}d show the conduction bands and associated spin texture around the M point, respectively. The spin splitted bands are predominately linear in \textbf{\textit{k}}. For deeper insights, the band dispersion around the high-symmetry-point can de deduced by identifying all the symmetry allowed terms such that $H(\textbf{\textit{k}})=O^\dagger H(\textbf{\textit{k}}) O$, where $O$ is symmetry operations belonging to the group of wave vector ($G$) associated with the high-symmetry-point and time-reversal symmetry ($T$)~\cite{voon2009kp}. The invariant Hamiltonian should satisfy the condition given below
 \begin{equation}
 	H_G(\textbf{\textit{k}})= D(O)H(O^{-1}\textbf{\textit{k}})D^{-1}(O),  \quad \forall O \in G, T
 \end{equation}
 where $D(O)$ is the matrix representation of operation $O$ belonging to point group $G$. The \textbf{\textit{k.p}} Hamiltonian is derived using the method of invariants by considering the little  group of M point to be $C_s$, consisting of mirror planes $M_{xz}$ and $M_{xy}$ besides identity operation ($E$). The \textbf{\textit{k.p}} Hamiltonian around the M point following the transformation rules listed in Table~\ref{tran} is given by
 \begin{equation}
 	H_{C_s}(\textbf{\textit{k}})=H_0(\textit{\textbf{k}})+\gamma k_y \sigma_z + \delta k_z \sigma_y
 	\label{e_1}
  \end{equation} 
where $H_0(\textit{\textbf{k}})$ is the part of Hamiltonian describing the band dispersion, depending on the parameter $\alpha$  and $\beta$ as:
\begin{equation}
	H_0(\textit{\textbf{k}})= E_0 + \alpha k_x^2+\beta k_y^2
\end{equation}
$\alpha$ and $\beta$ are related to the effective masses ($m_x^*$, $m_y^*$ ) by the expressions $\lvert\alpha\rvert$ = $\frac{\hslash^2}{2m_x^*}$ and $\lvert\beta\rvert$ = $\frac{\hslash^2}{2 m_y^*}$, respectively. $\sigma$ are the Pauli matrices describing spin degrees of freedom. $\gamma$ and $\delta$ are the linear SOC splitting coefficients. For dispersion relation in ($k_x-k_y$) plane around M point, Eq.~\ref{e_1} takes the form
\begin{equation}
	H_{C_s}(\textbf{\textit{k}})=H_0(\textit{\textbf{k}})+\gamma k_y \sigma_z 
	\label{e_M}
\end{equation}
The eigenstates corresponding to Eq.~\ref{e_M} are given by
\begin{equation}
	\Psi_{\textbf{\textit{k}}\uparrow}=e^{i\textbf{\textit{k}}_{\uparrow}.r}\begin{pmatrix}
		1  \\
		0 
	\end{pmatrix}
\end{equation}
and
\begin{equation}
	\Psi_{\textbf{\textit{k}}\downarrow}=e^{i\textbf{\textit{k}}_{\downarrow}.r}\begin{pmatrix}
		0  \\
		1 
	\end{pmatrix}
\end{equation}
The corresponding eigenvalues given by 
\begin{equation}
	E^{\pm}_{C_s}(\textbf{\textit{k}})=E_0 + \alpha k_x^2+\beta k_y^2\pm\gamma k_y
\end{equation}
The spin textures are determined by the expression $S_{\pm}=\frac{\hslash}{2} \langle \Psi_{\textbf{\textit{k}}\uparrow,\downarrow}\lvert \sigma\rvert\Psi_{\textbf{\textit{k}}\uparrow,\downarrow} \rangle$ are given by
\begin{equation}
	S_{\pm} = \pm \frac{\hslash}{2} (0,0,1)
	\label{e_2}
\end{equation}
Eq.~\ref{e_2} shows that only out-of-plane spin component is present around M point and found to be consistent with DFT spin texture shown in Figure~\ref{fig_band}d. By fitting the DFT calculated band structure and spin texture around the M point, we find that $\gamma=3.1$ eV\AA. This splitting is significantly larger as compared to other bulk systems such as BiTeCl ($\alpha_R=1.2$ eV\AA)~\cite{landolt2013bulk}, BiAlO$_3$ ($\alpha_R=0.74$ eV\AA)~\cite{da2016rashba} and LaWN$_3$ ($\alpha_R=2.7$ eV\AA)~\cite{zhao2020large,bandyopadhyay2020origin}. 

Figure~\ref{fig_band}e shows the splitting around $\Gamma$ point. The spin splitted bands around $\Gamma$ are purely cubic in \textbf{\textit{k}} with the only $z$-component is non vanishing (see Figure~\ref{fig_band}b). The splittings around the $\Gamma$ points can be understood in term of effective \textbf{\textit{k.p}} Hamiltonian. Here, group of the wave vector associated with the $\Gamma$  point is $D_{3h}$, comprising of trivial identity operation ($E$), three-fold rotation operation ($C_3$) and four mirror planes ($M_{xy}$, $M_{xz}$, $M'_{xz}$ and $M''_{xz}$). The Hamiltonian for ($k_x-k_y$) plane taking into account the symmetry invariants upto cubic in \textbf{\textit{k}} (see Table~\ref{tran}) can be expressed as
 \begin{equation}
 	H_{D_{3h}}(\textit{\textbf{k}})= H_0(\textit{\textbf{k}})+\lambda k_y(3k_x^2-k_y^2)\sigma_z
 	\label{eq_6m2}
 \end{equation}
where, $\lambda$ is the cubic SOC splitting coefficient. The eigenstates corresponding to this Hamiltonian are $\Psi_{k\downarrow}$ and $\Psi_{k\uparrow}$. The eigenvalues of Equation~\ref{eq_6m2}, which are splitted cubically, are obtained as
\begin{equation}
  E^{\pm}_{D_{3h}}(\textbf{\textit{k}}) = E_0 + \alpha k_x^2+ \beta k_y^2 \pm  3\lambda k_x^2k_y \mp \lambda k_y^3
\end{equation}
  The expectation values for the spin operators are same as given by Equation~\ref{e_2}, confirming the unidirectional spin texture around $\Gamma$ point. Note that Hamiltonian given by Equation~\ref{eq_6m2} depends upon the choice of coordinate system and valid only for space groups $P\overline{6}m2$ and $P\overline{6}c2$. The symmetry allowed term for $D_{3h}$ point group associated with $P\overline{6}2m$ and $P\overline{6}2c$ is $\zeta k_x(k_x^2-3k_y^2)\sigma_z$ instead of $\lambda k_y(3k_x^2-k_y^2)\sigma_z$. This is particularly due to the fact that mirror operation present in  $P\overline{6}2m$ and $P\overline{6}m2$ is $M_{yz}$ instead of $M_{xz}$. Additionally, $P\overline{6}$ space group is linked to the $C_{3h}$ point group and at $\Gamma$ point, little group remains $C_{3h}$. The effective two-band Hamiltonian, upto the third order in \textit{\textbf{k}} in the ($k_x-k_y$) plane is given by~\cite{zhao2020purely}
  \begin{equation}
  	H_{C_{3h}}(\textit{\textbf{k}})= H_0(\textit{\textbf{k}})+ \zeta k_x(k_x^2-3k_y^2)\sigma_z + \lambda k_y(3k_x^2-k_y^2)\sigma_z
  	\label{6}
  \end{equation}
and  the corresponding energy eigenvalues with eigenstates ($\Psi_{\textbf{k}\uparrow}$, $\Psi_{\textbf{k}\downarrow}$)
\begin{equation}
	 E^{\pm}_{C_{3h}}(\textit{\textbf{k}}) = E_0 + \alpha k_x^2+ \beta k_y^2 \pm \zeta k_x^3 \pm 3\lambda k_x^2k_y \mp 3\zeta k_y^2k_x \mp \lambda k_y^3
\end{equation}
The additional cubic splitting term with coefficient $\zeta$ arises for $C_{3h}$ point group, because the symmetry  is reduced as compared to $D_{3h}$ point group. Around the $\Gamma$ and M points, the bloch states ($\Psi_{\textbf{k}\uparrow}$, $\Psi_{\textbf{k}\downarrow}$) are the eigenstates of $\sigma_z$, leading to zero in-plane spin polarization. The only out-of-plane spin polarization is non-vanishing leading to PST  over full $\Gamma-$M$-$K plane in $P\overline{6}$, $P\overline{6}m2$, $P\overline{6}2m$, $P\overline{6}2c$ and $P\overline{6}c2$ space group materials. Splitting observed around M and $\Gamma$ points are linear and cubic in nature, respectively. The \textit{\textbf{k.p}} model and DFT give the same nature of spin splitting and textures. It is also worth emphasizing that the above two-bands models are also applicable to the other \textbf{\textit{k}}-points, whose little group is identical to the $\Gamma$ or M points. In  $P\overline{6}$, $P\overline{6}m2$, $P\overline{6}2m$, $P\overline{6}2c$ and $P\overline{6}c2$ space groups, the little group of A and L points are same as for $\Gamma$ and M points, respectively. Thus, the PST over full L$-$A$-$H with PCS around A point is also observed (see Figure~\ref{fig_band}f).
\begin{figure}[t]
	\begin{center}
		\includegraphics[width=17cm]{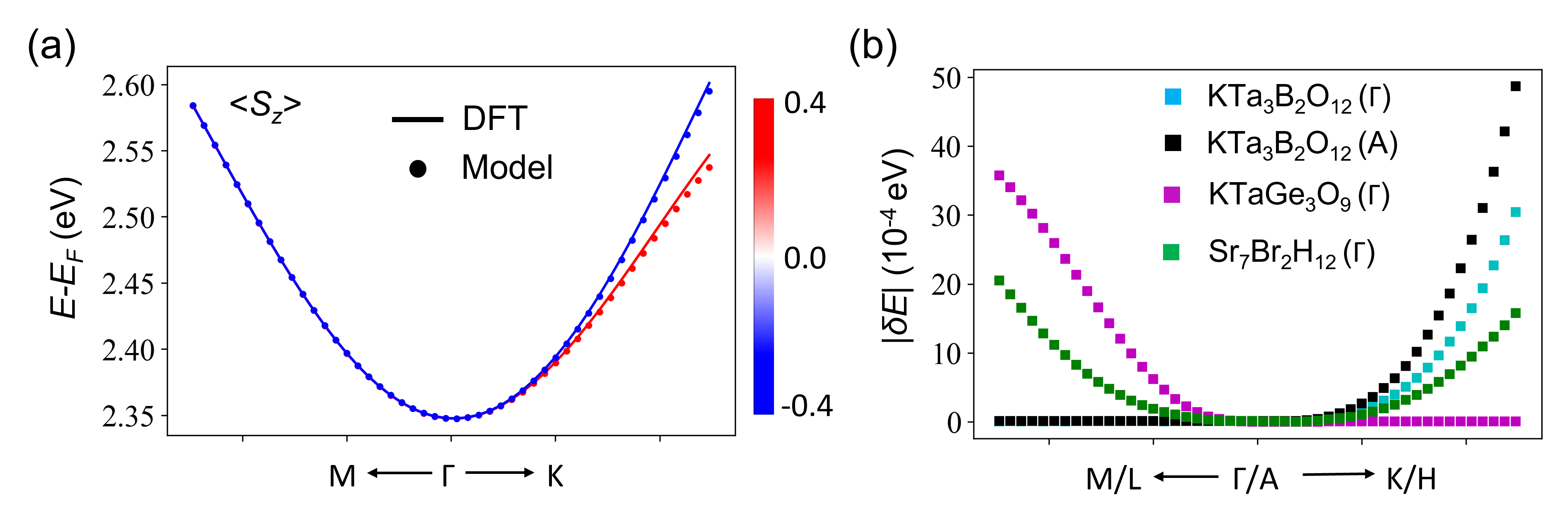}
		\caption{(a) Conduction bands of K$_3$Ta$_3$B$_2$O$_{12}$ around the $\Gamma$ point along M$-\Gamma-$K path projected over $z$-component of spin direction. (b) Spin-splitting energies ($\delta E$) of K$_3$Ta$_3$B$_2$O$_{12}$ (CBm), KTaGe$_3$O$_{9}$ (VBM) and Sr$_7$Br$_2$H$_{12}$ (VBM) are shown along high-symmetry-path M$-\Gamma-$K or L$-$A$-$H. The magnitude of  $\delta E$, is defined as $|\delta E|=|E(k,\uparrow)- (k,\downarrow)|$, where $E(k,\uparrow)$ and $E(k,\downarrow)$ are the energy bands with up and down spins, respectively.}
		\label{fig2}
	\end{center}
\end{figure}
 \begin{figure}[h]
	\begin{center}
		\includegraphics[width=15cm]{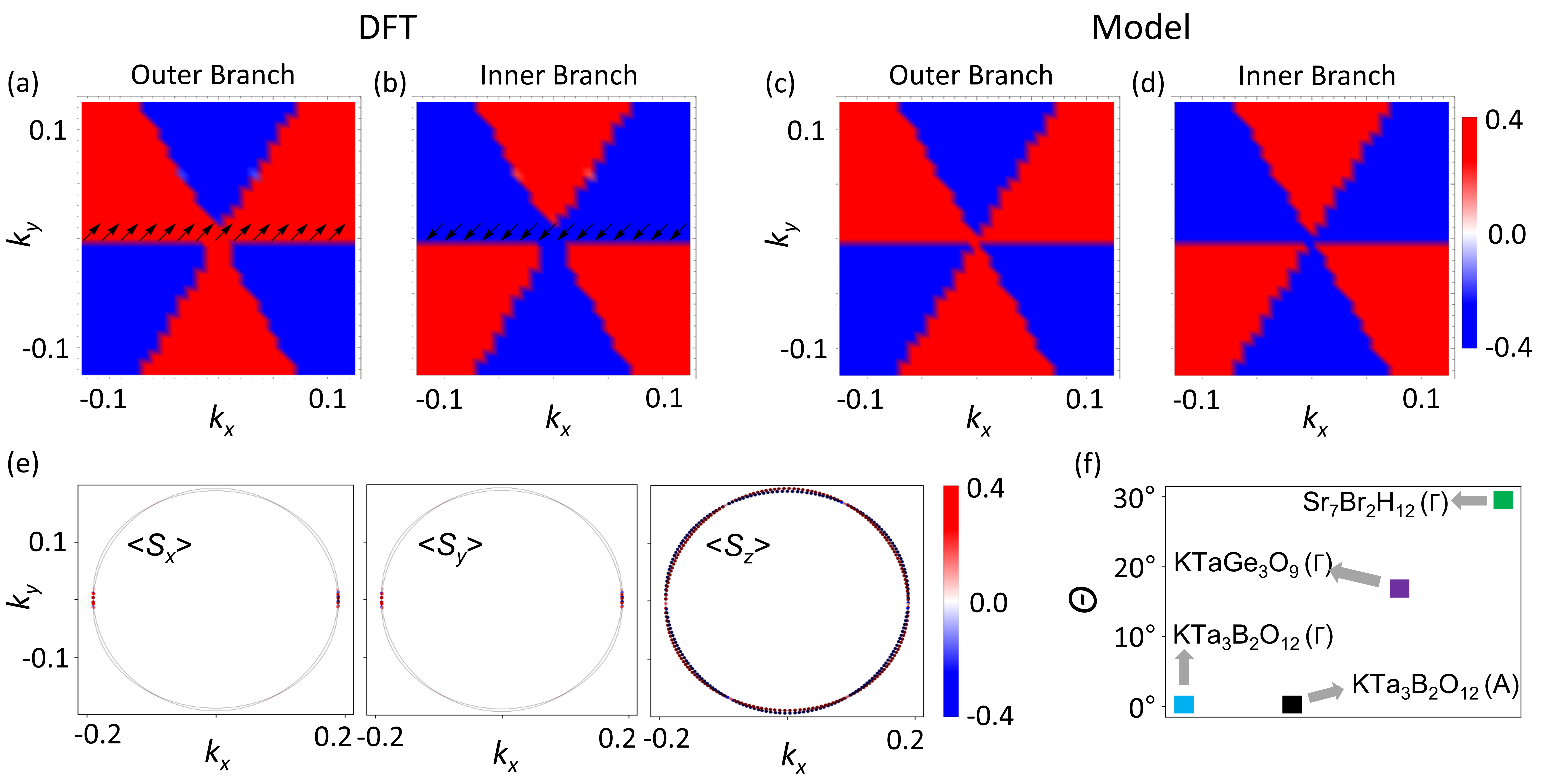}
		\caption{Spin textures of the conduction bands for K$_3$Ta$_3$B$_2$O$_{12}$ around the $\Gamma$ point obtained by DFT [(a)-(b)] and \textbf{\textit{k.p}} model [(c)-(d)] . (e) Constant energy contours projected over the spin components having energy of $E_F+2.38$ eV. (f) The calculated values of $\Theta$ for  spin textures of K$_3$Ta$_3$B$_2$O$_{12}$, KTaGe$_3$O$_{9}$ and Sr$_7$Br$_2$H$_{12}$.}
		\label{fig3}
	\end{center}
\end{figure} 

We have compared the band structures and spin textures of K$_3$Ta$_3$B$_2$O$_{12}$ around $\Gamma$ point with HSE06, PBE, PBEsol and PBE+U (here, U is the Hubbard parameter) $\epsilon_{xc}$ functionals to validate the choice of functional. The value of U is taken to be 2.5 eV to include the on-site Coulomb repulsion energy for Ta-5d orbitals, in accordance with previous studies~\cite{pasquier2022ab}. It is found that all the spin splitting properties except band gap calculated with HSE06, PBE, PBE+U are within $\pm$5\% as compared to PBEsol (see section V of SI). Our calculations have very little to do with the chosen functional, thus are based on computationally efficient PBEsol functional.  Firstly, we have calculated the band structure for the splitting bands along $k_x$ and $k_y$ directions. Then, we have fitted the band structure to \textit{\textbf{k.p}} model and the results are shown in Figure~\ref{fig2}a for K$_3$Ta$_3$B$_2$O$_{12}$. Our model fits the DFT band structure very well. The observed splitting is found to be highly anisotropic and depends upon the space group symmetry of the material. The bands are spin splitted along $\Gamma-$K direction and degenerate along $\Gamma-$M direction. Anisotropic effective mass could lead to the anisotropic splitting, but in our case, the effective mass is observed to be isotropic in both $\Gamma-$M and $\Gamma-$K directions. To further analyze the anisotropy, we have plotted the magnitude of the energy difference ($|\delta E|=|E(k,\uparrow)- (k,\downarrow)|$) between the splitted bands (see Figure ~\ref{fig2}b). The splitting is completely absent in $\Gamma-$M and A-L  directions for K$_3$Ta$_3$B$_2$O$_{12}$  and $\Gamma-$K direction for KTaGe$_3$O$_{9}$. The splitting for Sr$_7$Br$_2$H$_{12}$ is present along both  $\Gamma-$M and $\Gamma-$K directions with different magnitude. The different contributions coming from the terms with splitting coefficients $\zeta$ and $\lambda$ lead to anisotropic splitting along $\Gamma-$M and $\Gamma-$K directions.

Furthermore, the spin textures around the $\Gamma$ point and A point for the considered systems are also obtained using DFT and \textbf{\textit{k.p}} models. Figure~\ref{fig3}(a)-(b) are the obtained spin textures of K$_3$Ta$_3$B$_2$O$_{12}$ around $\Gamma$ point. The spin textures are momentum (\textbf{\textit{k}}) independent with non-zero out-of-plane components with a mere in-plane component around $\Gamma$ point in K$_3$Ta$_3$B$_2$O$_{12}$. Figure~\ref{fig3}(c)-(d) shows the spin textures calculated using \textbf{\textit{k.p}} model and are in accordance with the DFT results. It is interesting to note that the spin textures of considered configurations are different from each other. The border lines separating the $+|s_z|$ and $-|s_z|$ sectors vary case by case. In fact the border lines are the directions along which the energies of two bands become equal (see spin-projected constant energy contours in Figure~\ref{fig3}(e)). We introduce the parameter $\Theta$ to distinguish the spin textures as shown in Figure~\ref{fig1}b. These directions can be estimated from model by evaluating the $ \Theta= tan^{-1}(k_y/k_x)$ with $k_x \ge 0 $ and given by
\newcommand{\threepartdef}[6]
{
	\left\{
	\begin{array}{lll}
		#1  \\
		#2  \\		#3 
	\end{array}
	\right.
}
\begin{equation}
	tan(\Theta) = \threepartdef
	{-\frac{\zeta}{\lambda}-\frac{(\zeta^2+\lambda^2)^{2/3}}{\lambda (\zeta+i|\lambda|)^{1/3}}-\frac{(\zeta^2+\lambda^2)^{1/3}(\zeta+i|\lambda|)^{1/3}}{\lambda }}      {-\frac{\zeta}{\lambda}+\frac{(1-i\sqrt{3})[(\zeta^2+\lambda^2)^{1/3}(\zeta+i|\lambda|)^{1/3}]}{2\lambda}+\frac{(1+i\sqrt{3})(\zeta^2+\lambda^2)^{2/3}}{2\lambda (\zeta+i|\lambda|)^{1/3}}}
	{-\frac{\zeta}{\lambda}+\frac{(1+i\sqrt{3})[(\zeta^2+\lambda^2)^{1/3}(\zeta+i|\lambda|)^{1/3}]}{2\lambda}+\frac{(1-i\sqrt{3})(\zeta^2+\lambda^2)^{2/3}}{2\lambda (\zeta+i|\lambda|)^{1/3}}}
	{0}      {a^2 ,|, n \mbox{ for some } a > 1}
	{(-1)^r} 
\end{equation}
Figure~\ref{fig3}(f) shows $\Theta$ for the considered configurations. It is found to be consistent with DFT results (see section VI of SI). The angular difference between any two border lines of spin textures is calculated to be $60^{\circ}$ leading to the threefold rotation symmetry, which is in line with the threefold rotation symmetry of the crystal. Our calculations show that there is rather a big class of materials that shows PST with PCS. All of these materials along with the calculated parameters are reported in Table~\ref{tbl1}. Table~\ref{tbl1} also includes the already reported parameters of Ge$_{3}$Pb$_{5}$O$_{11}$, Pb$_{7}$Cl$_{2}$F$_{12}$ and Pb$_{7}$Br$_{2}$F$_{12}$ for comparison. The splitting observed around $\Gamma$ point for K$_3$Ta$_3$B$_2$O$_{12}$ ($\lambda=6.85$eV\AA$^3$) is larger as compared to Ge$_{3}$Pb$_{5}$O$_{11}$ ($\chi = -5.24$ eV\AA$^3$), Pb$_{7}$Br$_{2}$F$_{12}$ ($\lambda = -5.04$ eV\AA$^3$) and comparable to Pb$_{7}$Cl$_{2}$F$_{12}$ ($\lambda = 8.65$ eV\AA$^3$). Splitting around A point of K$_3$Ta$_3$B$_2$O$_{12}$ is $\lambda=8.13$eV\AA$^3$, larger than that of around $\Gamma$ point. The magnitude of splitting in these materials can be further increased by doping of heavier elements with large SOC~\cite{volobuev2017giant} and introducing strain~\cite{arras2019rashba,bhumla2021origin} to make its experimental detection more easily accessible (i.e. spin-resolved photoemission spectroscopy).
\begin{figure}
	\begin{center}
		\includegraphics[width=12cm]{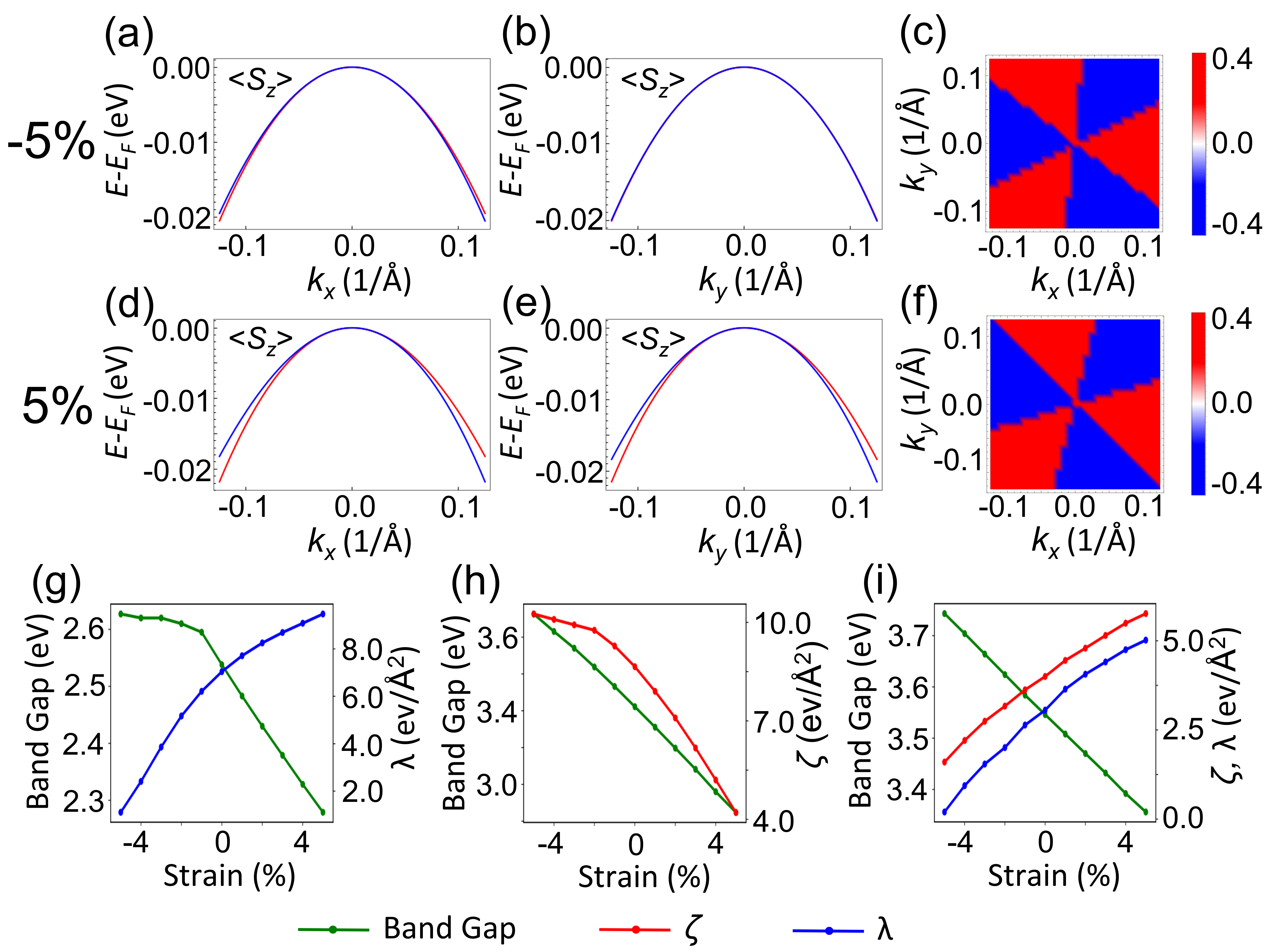}
		\caption{ Band structures and spin texture of the valence bands for Sr$_{7}$Br$_{2}$H$_{12}$ around the $\Gamma$ point under $-5$\% [(a)-(c)] and $5$\% [(d)-(f)] strain obtained by DFT. Note that there should be two spin-texture branches associated to two bands; only one branch (associated to outer band) is shown for each case. The variation in band gap  and splitting parameters ($\zeta$ and $\lambda$) as a function of strain for (g) K$_3$Ta$_3$B$_2$O$_{12}$, (h) KTaGe$_3$O$_{9}$ and (i) Sr$_7$Br$_2$H$_{12}$.  }
		\label{fig4}
	\end{center}
\end{figure}
\subsection*{Effects of strain}
Strain can affect the band structure of semiconductors by tuning band gaps, varying effective mass of carriers, shifting bands and introducing new band splittings~\cite{voon2009kp}. Tuning of bands gaps, varying effective mass of the carriers and shifting of bands are general consequences of strain. However new band splittings happens when the symmetry of a crystal reduces with the strain~\cite{skierkowski2007strain}. Effect of the strain is extensively studied for LR and LD effects~\cite{arras2019rashba,bhumla2021origin,tao2016strain,leppert2016electric, absor2016strain,sheoran2021rashba, anshory2020strain}, but effect of strain on PST with PCS is still unknown. Here, we have introduced the out-of-plane strain such that the point group symmetry of crystal remains intact. In case of materials with $D_{3h}$ point group, the border lines separating the $+|s_z|$ and $-|s_z|$ sectors ($\Theta$) remain same. Therefore, PSTs including border line directions are robust to the strain and protected by the symmetry. Only splitting coefficients for K$_3$Ta$_3$B$_2$O$_{12}$ and KTaGe$_3$O$_{9}$ can be tuned with the strain. In contrast, strain tunes the spin splitting as well as spin textures for Sr$_7$Br$_2$H$_{12}$. We have varied the strain from $-5$\% to $5$\%, where "$-$" and "$+$" denote the compressive and tensile strain, respectively. The phonon spectra show that  Sr$_7$Br$_2$H$_{12}$ is dynamically stable under such strain (see section VII of SI). Figure~\ref{fig4}(a)-(f) show the splitted bands and corresponding spin textures under the $\pm 5\%$ strain for Sr$_7$Br$_2$H$_{12}$. Splitting increases (decreases) under tensile (compressive) strain. Figure~\ref{fig4}(g)-(i) shows the variation of band gaps and splitting coefficients for K$_3$Ta$_3$B$_2$O$_{12}$, KTaGe$_3$O$_{9}$ and Sr$_7$Br$_2$H$_{12}$ and  are sensitive to strain. The trend observed is linear same as for most of the linear splitting. In case of K$_3$Ta$_3$B$_2$O$_{12}$, the splitting can be increased from $6.85$ eV\AA$^3$ (without any strain) to $9.45$ eV\AA$^3$ at $+5\%$ strain.
\section*{Conclusions}
Combining first principles calculations and symmetry analysis, we have shown the existence of PST in bulk materials having $D_{3h}$ and $C_{3h}$ point group symmetry. The unidirectional out-of-plane spin textures are observed in full $\Gamma-$M$-$K plane, as long in-plane mirror symmetry remains intact and differs from trivial PST. In-plane mirror symmetry along with the three-fold rotation symmetry leads to purely cubic spin splitting. Taking K$_3$Ta$_3$B$_2$O$_{12}$ as a test case or prototype, we have observed purely cubic splitting around the $\Gamma$ point of order $\approx6.85$ eV\AA$^3$, which is larger than already reported splitting. Around M point, bands are linear splitted with order of $\approx3.1$ eV\AA. The observed splitting are found to be anisotropic in nature. The nature and anisotropy of splitting are also studied using the \textbf{\textit{k.p}} model via symmetry analysis. Our calculations show that there exist a large family of materials which harbour full plane PST in purely cubical regime. In addition, strain engineering tunes the observed PST by varying the magnitudes of SOC splitting coefficients. These spin textures known for non-dissipative spin transport, together with PCS form another prospective aspect. The large SOC splitting coefficients, wide band gap, suitable band edge positions, strain tunability and room temperature stability make them suitable for room temperature applications. The complete realization of original device of these applications may enrich the field of spintronics.
\begin{table}[h]
	\begin{center}
		\caption{ The materials showing PCS with PST along with their parameters of Equation~\ref{6}. The splitting is observed for low conduction bands (CB) or high valence bands (VB) around the high-symmetry points (HSP) $\Gamma$ and A.}
		\label{tbl1}
		
		\begin{tabular}{ p{2.0 cm} p{1.7cm} p{1.7cm} p{1.7cm} p{1.7cm}  p{1.7cm}  p{1.7cm} p{2.0cm} }
			\hline
			\hline
			Materials     &  CB/VB   & HSP & ${E_0}$ (eV)& {$\alpha$} (eV\AA$^2$) & {$\zeta$} (eV\AA$^3$) & {$\lambda$} (eV\AA$^3$) & {Ref} \\ \hline \hline
			BaHfSi$_{3}$O$_{9}$ & VB & $\Gamma$ & -0.19 & 3.05 &  0.65 & 0.00 &\\
			BiTa$_{7}$O$_{19}$   & VB & $\Gamma$ & 0.00 & -3.79 & 2.35 & 0.00& \\
			BiTa$_{7}$O$_{19}$    & CB & $\Gamma$ & 2.82 & -1.20     & 0.55 & 0.00& \\
			LaTa$_{7}$O$_{19}$   & VB & $\Gamma$ & -0.19 & -3.28   & 0.95 & 0.00& \\
			LaTa$_{7}$O$_{19}$   & CB & $\Gamma$ & 3.56  & -1.23     & 9.55 & 0.00& \\
			KCaP$_{3}$O$_{9}$   & VB & $\Gamma$ & -0.15  & 1.96  & 0.85 & 0.00 &\\
			KMgP$_{3}$O$_{9}$  & VB & $\Gamma$ & -0.19  & 2.13  & 1.00 & 0.00 &\\
			KTaGe$_{3}$O$_{9}$  & VB & $\Gamma$  & -0.35 & 1.44  & 8.05     & 0.00   &  \\
			RbNbGe$_{3}$O$_{9}$ & VB & $\Gamma$  & -0.11  & 1.26 &  4.30 & 0.00 & \\
			TlTaGe$_{3}$O$_{9}$ & VB & $\Gamma$  & -0.15  & 1.59 & 6.35 & 0.00 & \\
			TlTaGe$_{3}$O$_{9}$  & CB & $\Gamma$  & 3.58  & 1.86 & 1.52 & 0.00 & This work \\
			K$_{3}$Ta$_{3}$B$_{2}$O$_{12}$   & CB & $\Gamma$  & 2.34   &  5.04    & 0.00 & 6.85  &  \\
			K$_{3}$Ta$_{3}$B$_{2}$O$_{12}$       & CB & A   & 5.37    & 5.05   & 0.00 & 8.13   & \\
			K$_{4}$Au$_{6}$S$_{5}$   & VB & $\Gamma$   & 0.00   & -0.95 & 0.00 & 5.55 &\\
			K$_{4}$Au$_{6}$S$_{5}$        & CB & $\Gamma$   & 1.62   & 8.07 & 0.00 & 2.01 &\\
			RbS     & VB & $\Gamma$   &  0.00  & -0.91 & 0.00 & 1.06 &\\
			W$_{6}$CCl$_{18}$    & CB & $\Gamma$   & 1.04   & 0.51 & 0.00 & 9.10 &\\
			Sr$_{7}$Cl$_{2}$H$_{12}$   & VB & $\Gamma$   & 0.00   & -1.13     & 2.35     &  1.80   & \\
			Sr$_{7}$Br$_{2}$H$_{12}$  & VB   &  $\Gamma$ & 0.00  & -1.07   &  3.21    &  2.45   & \\
			Pb$_{7}$Br$_{2}$F$_{12}$  & CB  & A &  3.74& -3.11&-0.91 & -4.36 &\\
			Pb$_{7}$Cl$_{2}$F$_{12}$   & CB  & A & 3.98 & 5.11 &1.23& 8.18 &\\ \hline
			Pb$_{7}$Br$_{2}$F$_{12}$   & CB  & $\Gamma$ &  3.76& -1.11&-2.92 & -5.04 &\\
			Pb$_{7}$Cl$_{2}$F$_{12}$  & CB  & $\Gamma$ & 3.98   &-2.08 &1.75& 8.65 & \cite{zhao2020purely}\\
			Ge$_{3}$Pb$_{5}$O$_{11}$   & CB  & $\Gamma$& 2.13 &1.27 & -5.24& -3.43 & \\ \hline\hline
		\end{tabular}
	\end{center}
\end{table}
\section*{Methods}
We have employed the relativistic first-principles calculations based on density functional theory (DFT) as implemented in Vienna \textit{Ab initio} Simulation Package (VASP)~\cite{kresse1993ab, kresse1996efficiency}. The simulations are done using plane-wave basis set and projector augmented wave method~\cite{kresse1994norm,kresse1999ultrasoft}. We have used the Perdew-Burke-Ernzerhof revised for solids (PBEsol) as the exchange-correlation ($\epsilon_{xc}$) functional~\cite{perdew2008restoring} and set plane-wave cut off energy to 550 eV. To get a more accurate band gap, calculations are also performed using non-local Heyd-Scuseria-Ernzerhof (HSE06) $\epsilon_{xc}$ functional~\cite{heyd2003hybrid}.  Initially, experimental lattice parameters and atomic positions are taken as a starting point. In structural optimization, the change in total energy between two electronic steps is set to 10$^{-6}$ eV and are converged until Hellmann-Feynman forces are smaller than 1 meV\AA$^{-1}$ without including spin-orbit coupling. The Brillouin zone was sampled in \textit{\textbf{k}}-space with Monkhorst-Pack~\cite{monkhorst1976special} scheme with a \textit{\textbf{k}}-mesh of 6$\times$6$\times$12 for K$_3$Ta$_3$B$_2$O$_{12}$, KTaGe$_3$O$_{9}$ and 8$\times$8$\times$6 for Sr$_7$Br$_2$H$_{12}$. After the structure optimization, we have also confirmed the dynamical stability using density functional perturbation theory (DFPT). The phonon dispersion curves are calculated by considering 2$\times$2$\times$2 supercell with the PHONOPY code~\cite{togo2008first}. The spin textures from DFT are calculated using a dense \textit{\textbf{k}}-mesh of 51$\times$51 around high symmetry point in $k_x-k_y$ plane. The model band structures and spin textures are calculated by parameterizing the models using minimization of the summation
\begin{equation}
\begin{split}
S=\sum_{i=1}^{2} \sum_{k_x, k_y, k_z}^{} f(k_x, k_y, k_z) |Det[H(k_x, k_y, k_z)-E^{i}(k_x, k_y, k_z)I]|^2
\end{split}
\label{Eq_2}
\end{equation}
over i$^{th}$ energy eigenvalues ($E^i(k_x,k_y,k_z)$) as training sets, where f($k_x,k_y, k_z$) is the weight attached to ($k_x,k_y, k_z$) point.  Here, H, "Det" and $I$ represent model Hamiltonian, determinant and identity matrix, respectively. We have used normal distribution for f($k_x,k_y, k_z$) centered at ($k_x,k_y, k_z$) point to get a better fit near high-symmetry point as used in Ref\cite{zhao2020large, zhao2020purely}. We have conducted symmetry analysis using Ref~\cite{koster1963properties}, Bilbao crystallographic server~\cite{aroyo2006bilbao, elcoro2017double}, SEEK-PATH software~\cite{hinuma2017band} and FINDSYM~\cite{stokes2005findsym}. The Mathematica~\cite{Mathematica} and PyProcar~\cite{herath2020pyprocar} are used to plot spin textures, band structures and parameterize the models. Since our considered materials are nonmagnetic, we have not artificially initialized spin configurations;
the final spin configurations are determined by VASP after fully
converging the electronic self-consistent loop. For considering the strain, we have varied the lattice parameter ($c$)  with respect to equilibrium lattice parameter ($c_o$) and further relaxed the atomic coordinates.

\bibliography{sci}



\section*{Acknowledgements}

S.S. acknowledges  CSIR,  India,  for  the  junior  research fellowship [Grant No. 09/086(1432)/2019-EMR-I]. P.B. acknowledges UGC, India, for the senior research fellowship [Grant No. 1392/(CSIR-UGC NET JUNE 2018)]. S.B. acknowledges the financial support from SERB under Core  Research  Grant [Grant  No.  CRG/2019/000647]. We acknowledge the High Performance Computing (HPC) facility at IIT Delhi for computational resources.

\section*{Author contributions statement}

S.S. and S.B. conceived the project. S.B. supervised overall. S.S. performed all the calculations. S.S. and P.B. got involved in various discussion to analyze the data. All authors took part in finalizing the manuscript.

\section*{Additional information}

\textbf{Competing interests:} The authors declare no competing interests.




\end{document}


\title{\textbf{\Large{Full-plane persistent spin textures with cubic order intrinsic and anisotropic band splitting in bulk Lead-free materials}}}
	\author{Sajjan Sheoran$^*$, Preeti Bhumla,  Saswata Bhattacharya$^*$\\Department of Physics, Indian Institute of Technology Delhi, New Delhi 110016 India \\
		$^*$Email: sajjan@physics.iitd.ac.in [SS], saswata@physics.iitd.ac.in [SB]}
	\pacs{}
	\maketitle
\begin{center}
	{\Large \bf Supplemental Material}\\ 
\end{center}
\begin{enumerate}[\bf I.]
	\item Experimentally synthesizable materials having $P\overline{6}2m$, $P\overline{6}2c$, $P\overline{6}m2$, $P\overline{6}c2$ and $P\overline{6}$ space group symmetry
	\item Crystal structures and atomic coordinates of K$_3$Ta$_3$B$_2$O$_{12}$, KTaGe$_3$O$_9$ and Sr$_7$Br$_2$H$_{12}$
	\item Stability analysis using density function perturbation theory (DFPT)
	\item Electronic band structures of K$_3$Ta$_3$B$_2$O$_{12}$, KTaGe$_3$O$_9$ and Sr$_7$Br$_2$H$_{12}$
	\item Comparison of HSE06, PBE, PBEsol and PBE+U exchange correlation functionals
	\item Spin textures
	\item Stability under strain
\end{enumerate}
\newpage
\section{E\MakeLowercase{xperimentally synthesizable materials having }\textrm{$P$}\MakeLowercase{$\overline{6}2\MakeLowercase{m}$}, \textrm{$P$}\MakeLowercase{$\overline{6}2\MakeLowercase{c}$}, \textrm{$P$}\MakeLowercase{$\overline{6}\MakeLowercase{m}2$}, \textrm{$P$}\MakeLowercase{$\overline{6}\MakeLowercase{c}2$ and} \textrm{$P$}\MakeLowercase{$\overline{6}$ space group symmetry}}
Among 230 space groups, four space groups $P\overline{6}2m$, $P\overline{6}2c$, $P\overline{6}m2$ and $P\overline{6}c2$ possess $D_{3h}$ point group symmetry. In addition, $P\overline{6}$ space group also contains $C_{3h}$ point group symmetry.  These five space groups may show persistent spin textures (PST) in the regime of purely cubic splitting (PCS). In view of this, we have chosen materials having these space groups. Table S1 shows the non-metallic and non-magnetic materials with $P\overline{6}2m$, $P\overline{6}2c$, $P\overline{6}m2$, $P\overline{6}c2$ and $P\overline{6}$ space group symmetries, which are already experimentally synthesized. Wide band gap materials have practical advantage of small leakage current for room tempertaure applications. Thus, we have included the materials with band gap greater than 1 eV. The space groups are calculated using FINDSYM~\cite{stokes2005findsym} software with tolerance of 0.001.

\begin{table}[h]
	\begin{center}
		\caption{Experimentally synthesizable non-metallic and non-magnetic materials with $P\overline{6}2m$, $P\overline{6}2c$, $P\overline{6}m2$, $P\overline{6}c2$ and $P\overline{6}$ space group symmetries. These materials are screened from the material database of AFLOW and Materials Project. The band gaps are obtained from the simulations based on DFT using PBEsol exchange correlation functional with inclusion of spin-orbit coupling (SOC), thus may vary from the band gaps reported at AFLOW or Materials Project~\cite{jain2013commentary}. ICSD-IDs show that the materials are already experimentally synthesized. The space groups are reported in accordance with FINDSYM software. }
		\label{tbl1}
		\begin{tabular}{p{4.2 cm} p{4.2cm} p{4.2cm} p{4.2cm}}
			\hline
			\hline
			{Materials}     &  {Space group}   & {ICSD-ID} & {Band Gaps (eV)}   \\ \hline
			K$_3$Ta$_3$B$_2$O$_{12}$  & $P\overline{6}2m$  &  201143  &  2.48   \\  
			Ba$_3$Ta$_6$Si$_4$O$_{26}$    & $P\overline{6}2m$  & 15935   & 2.82    \\
			RbS & $P\overline{6}2m$  &  73176  & 1.59   \\   
			KS  & $P\overline{6}2m$  &  73171  & 1.43   \\ 
			NaSrP  & $P\overline{6}2m$  & 416887   & 1.49   \\
			BaBPF$_{10}$  & $P\overline{6}2m$  &  420597  & 7.29   \\
			RbMgFCO$_3$  & $P\overline{6}2m$  &  238611  & 4.63   \\
			LaFCO$_3$  & $P\overline{6}2m$  & 26678   & 4.03   \\
			SrBe$_3$O$_4$  & $P\overline{6}2c$ & 26179   & 4.03   \\
			K$_4$Au$_6$S$_5$ & $P\overline{6}2c$  & 202552   & 1.62   \\
			W$_6$CCl$_{18}$  & $P\overline{6}2c$  &  413026  & 1.04   \\ 
			Li$_{10}$BrN$_3$  & $P\overline{6}m2$  & 78819   & 1.65   \\ 
			GaSe  & $P\overline{6}m2$  & 73387   & 0.48   \\ 
			YHSe  & $P\overline{6}m2$  &  72008  & 2.00   \\ 
			RbBe$_3$ZnF$_9$ & $P\overline{6}c2$  & 23133   & 6.16   \\ 
			KMgP$_3$O$_9$  & $P\overline{6}c2$  &  28012  & 4.88   \\ 
			KCaP$_3$O$_9$  & $P\overline{6}c2$  &  281588  & 4.98   \\ 
			BaHfSi$_3$O$_9$  & $P\overline{6}c2$  &  183835  & 4.57   \\ 
			TlTaGe$_3$O$_9$  & $P\overline{6}c2$  &  10383  & 3.58   \\ 
			KTaGe$_3$O$_9$ & $P\overline{6}c2$  &  10380  & 3.43   \\ 
			RbNbGe$_3$O$_9$  & $P\overline{6}c2$  &  10382  & 3.21   \\
			LaTa$_7$O$_{19}$  & $P\overline{6}c2$ & 60837   & 3.00   \\
			BiTa$_7$O$_{19}$  & $P\overline{6}c2$ & 251951   & 2.82   \\
			Sr$_7$Br$_2$H$_{12}$  & $P\overline{6}$  & 418949   & 3.52   \\
			Sr$_7$Cl$_2$H$_{12}$         & $P\overline{6}$ &  418948  & 3.68   \\
			Ca$_7$Cl$_2$H$_{12}$         & $P\overline{6}$  &  420927  & 3.70   \\
			Ba$_7$Cl$_2$F$_{12}$         & $P\overline{6}$  & 410679   & 5.95   \\
			Ga$_9$Tl$_3$S$_2$O$_{13}$         & $P\overline{6}$  & 61256   & 2.00   \\
			NaLiCO$_3$         & $P\overline{6}$ &  80459  & 4.33   \\
			BaZnBF$_3$        & $P\overline{6}$  & 429814   & 3.73   \\
			Ca$_{10}$P$_6$O$_{25}$         & $P\overline{6}$ & 87727   & 3.14   \\
			LiCdBO$_3$         & $P\overline{6}$ &  20191  & 2.00   \\
			CeCO$_4$        & $P\overline{6}$  &  238537  & 1.26  \\

			\hline\hline
			
		\end{tabular}
	\end{center}
\end{table}
\clearpage
\section{C\MakeLowercase{rystal structures and atomic coordinates of} K$_3$T\MakeLowercase{a}$_3$B$_2$O$_{12}$, KT\MakeLowercase{a}G\MakeLowercase{e}$_3$O$_{9}$ \MakeLowercase{and} S\MakeLowercase{r$_7$}B\MakeLowercase{r$_2$}H$_{12}$  }

Figure S1, S2 and S3 show the side and top views of the unit cell of K$_3$Ta$_3$B$_2$O$_{12}$, KTaGe$_3$O$_9$ and Sr$_7$Br$_2$H$_{12}$ with space groups $P\overline{6}2m$, $P\overline{6}c2$ and $P\overline{6}$, respectively. The unit cell of K$_3$Ta$_3$B$_2$O$_{12}$, KTaGe$_3$O$_9$ and Sr$_7$Br$_2$H$_{12}$ contains 20, 28 and 19 atoms, respectively. The wyckoff positions and structural coordinates optimized using PBEsol without SOC are given in Table S2. 

\begin{figure}[H]
	\includegraphics[width=18cm]{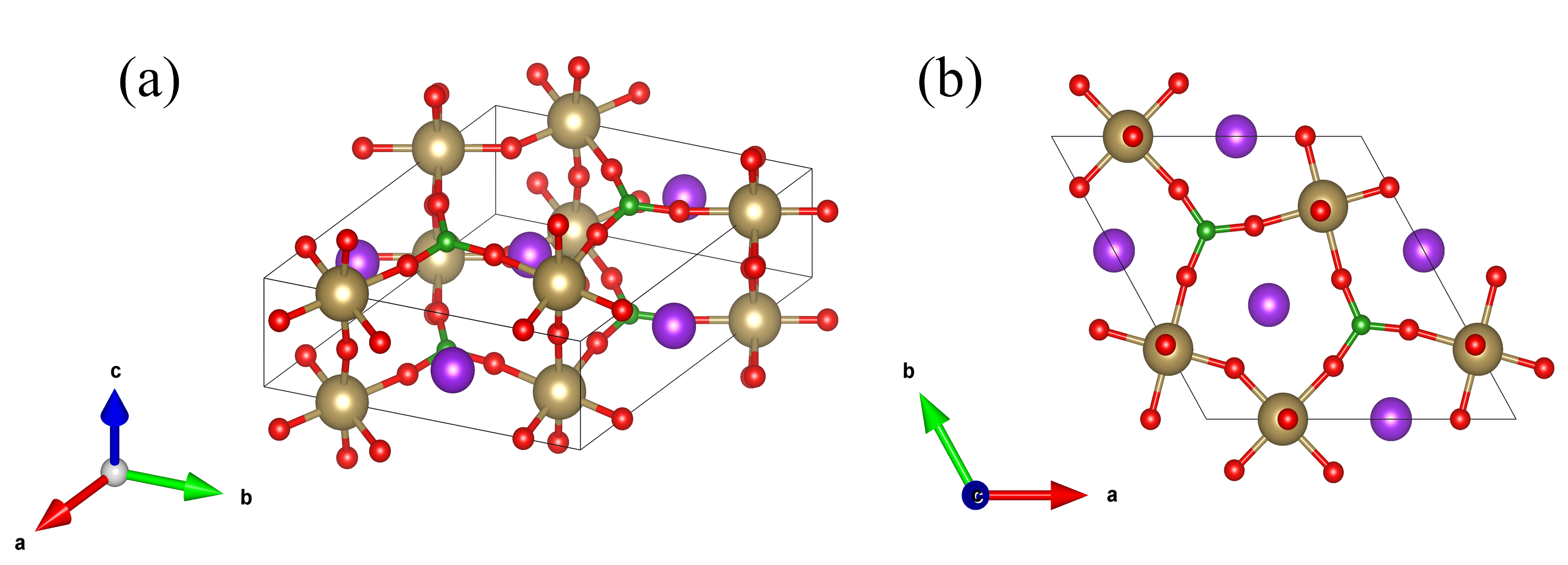}
	\caption{Crystal structure of K$_3$Ta$_3$B$_2$O$_{12}$ having $P\overline{6}2m$ space group symmetry with panels (a) and (b) being side and top views. The violet, grey, green and red balls denote K, Ta, B and O ions, respectively. }
\end{figure}
\begin{figure}[H]
	\includegraphics[width=18cm]{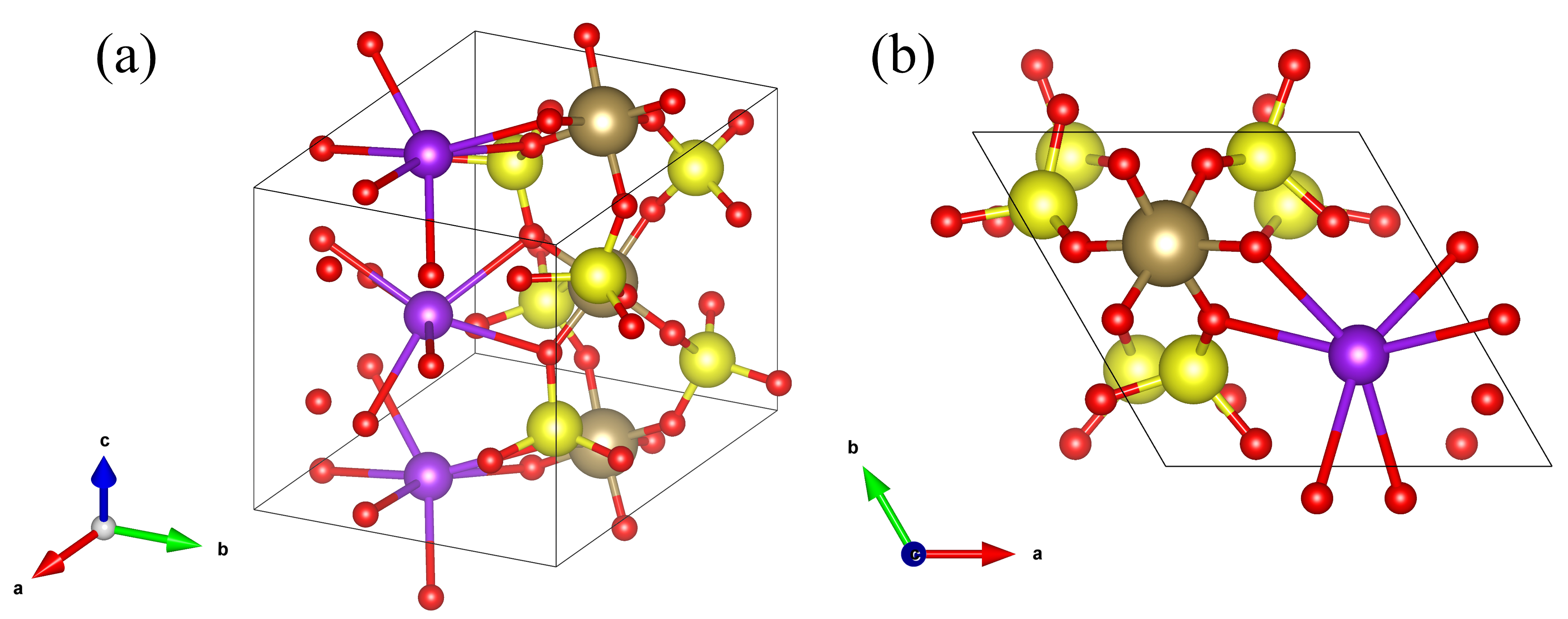}
	\caption{Crystal structure of KTaGe$_3$O$_9$ having $P\overline{6}c2$ space group symmetry with panels (a) and (b) being side and top views. The violet, grey, yellow and red balls denote K, Ta, Ge and O ions, respectively. }
\end{figure}
\begin{figure}[H]
	\includegraphics[width=18cm]{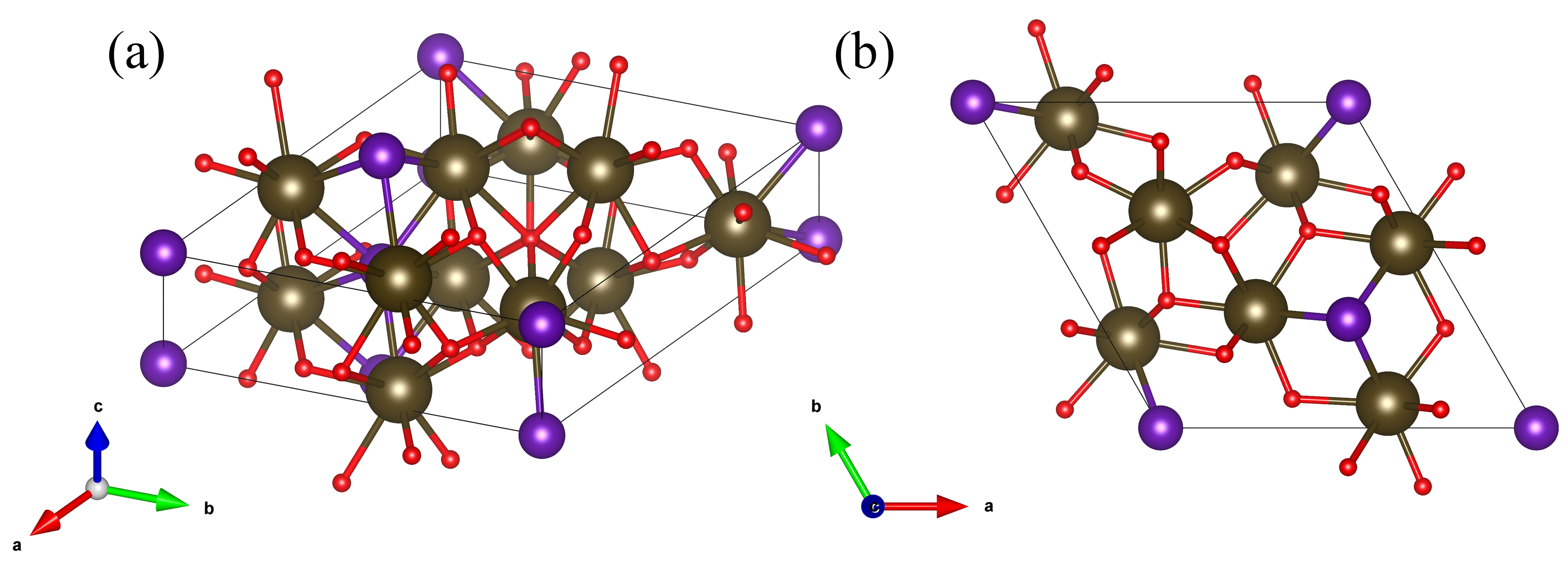}
	\caption{Crystal structure of Sr$_7$Br$_2$H$_{12}$ having $P\overline{6}$ space group symmetry with panels (a) and (b) being side and top views. The grey, violet and red balls denote Sr, Br and H ions, respectively. }
\end{figure}
\vfill
\vfill

\begin{table}[H]
	\begin{center}
		\caption{The DFT simulated atomic coordinates of K$_3$Ta$_3$B$_2$O$_{12}$ ($P\overline{6}2m$), KTaGe$_3$O$_9$ ($P\overline{6}c2$)and Sr$_7$Br$_2$H$_{12}$ ($P\overline{6}$). The space groups of various compounds are calculated using FINDSYM software.} 
		\label{tbl1}
		\begin{tabular}{ p{3.5cm} p{3.5cm} p{3.5cm} p{3.5cm} p{3.5cm} }
			\hline
			\hline
			& &  {K$_3$Ta$_3$B$_2$O$_{12}$} &      &     \\ 
			
			&  a = b = 8.88869 $\AA$  & c = 3.94721 $\AA$ & $\alpha$ = $\beta$ = 90$^\circ$   &  $\gamma$ = 120$^\circ$   \\ \hline
			{Ions}     &  {Wyckoff Position}   & {x}  & {y}   &  {z}   \\ 
			K & 3g  &0.59624538824000  & 0.00000190813178  & 0.50000000000000  \\  
			Ta & 3f & 0.24704722975237 &  0.00000177044865 &  0.00000000000000 \\
			B &  2c & 0.33332864241096 &  0.66666432120548 &  0.00000000000000 \\   
			O & 3g  &0.26282259961208  & 0.00000205345854  & 0.50000000000000  \\ 
			O & 3f  &0.81954320729084  & 0.99999987304096  & 0.00000000000000  \\
			O & 6j  &0.50415282187740  & 0.18790954220969  & 0.00000000000000  \\ 
			 \hline\hline
			& &  {KTaGe$_3$O$_{9}$} &      & \\
			&  a = b = 7.08419 $\AA$  & c = 10.29223 $\AA$ & $\alpha$ = $\beta$ = 90$^\circ$   &  $\gamma$ = 120$^\circ$   \\ \hline
			{Ions}     &  {Wyckoff Position}   & {x}  & {y}   &  {z}   \\ 
			K  & 2a  & 0.99997531080484  & 0.00002468919516 &  0.00000000000000  \\
			Ta & 2c  & 0.33333333333333  & 0.66666666666667 &  0.00000000000000  \\
			Ge & 6k  & 0.73969395626600  & 0.62326227939000 &  0.25000000000000  \\ 
			O  & 6k  & 0.93395178319055  & 0.53355333416771 &  0.25000000000000  \\
			O  & 12l & 0.76216331483556  & 0.77291907264606 &  0.10949977515197  \\
			\hline \hline 
			& & {Sr$_7$Br$_2$H$_{12}$}  &      &     \\ 
			&  a = b = 10.07849 $\AA$  & c = 4.02249 $\AA$ & $\alpha$ = $\beta$ = 90$^\circ$   &  $\gamma$ = 120$^\circ$   \\ \hline
			{Ions}     &  {Wyckoff Position}   & {x}  & {y}   &  {z} \\
			Sr& 1c & 0.33333711098159 &  0.66666822379182 &  0.00000000000000   \\
			Sr& 3j & 0.64180525906510 &  0.07451027729722 &  0.00000000000000   \\
			Sr& 3j & 0.05071770916156 &  0.27482880746735 &  0.50000000000000   \\ 
			Br& 1a & 0.00000000000000 &  0.00000000000000 &  0.00000000000000   \\
			Br& 1f & 0.66666419453259 &  0.33333165971454 &  0.50000000000000   \\
			H & 3j & 0.91051775568348 &  0.30563754378321 &  0.00000000000000   \\
			H & 3j & 0.21226069413806 &  0.39201165031733 &  0.00000000000000   \\
			H & 3k & 0.77326046722585 &  0.05643668310922 &  0.50000000000000   \\
			H & 3k & 0.12053233244245 &  0.55922308925490 &  0.50000000000000   \\
			\hline \hline
		\end{tabular}
	\end{center}
\end{table}
\section{S\MakeLowercase{tability analysis using density functional perturbation theory} (DFPT)}
The dynamical stability of K$_3$Ta$_3$B$_2$O$_{12}$, KTaGe$_3$O$_{9}$ and Sr$_7$Br$_2$H$_{12}$ is analyzed by plotting the phonon spectra. Phonon dispersion curves are calculated using DFPT approach in PHONOPY~\cite{phonopy} code, with a supercell of 2$\times$2$\times$2. Figure S4 shows the calculated band dispersion curves along high-symmetry-directions. The absence of negative frequencies confirms the stability at low temperatures.
\begin{figure}[H]
	\includegraphics[width=16cm]{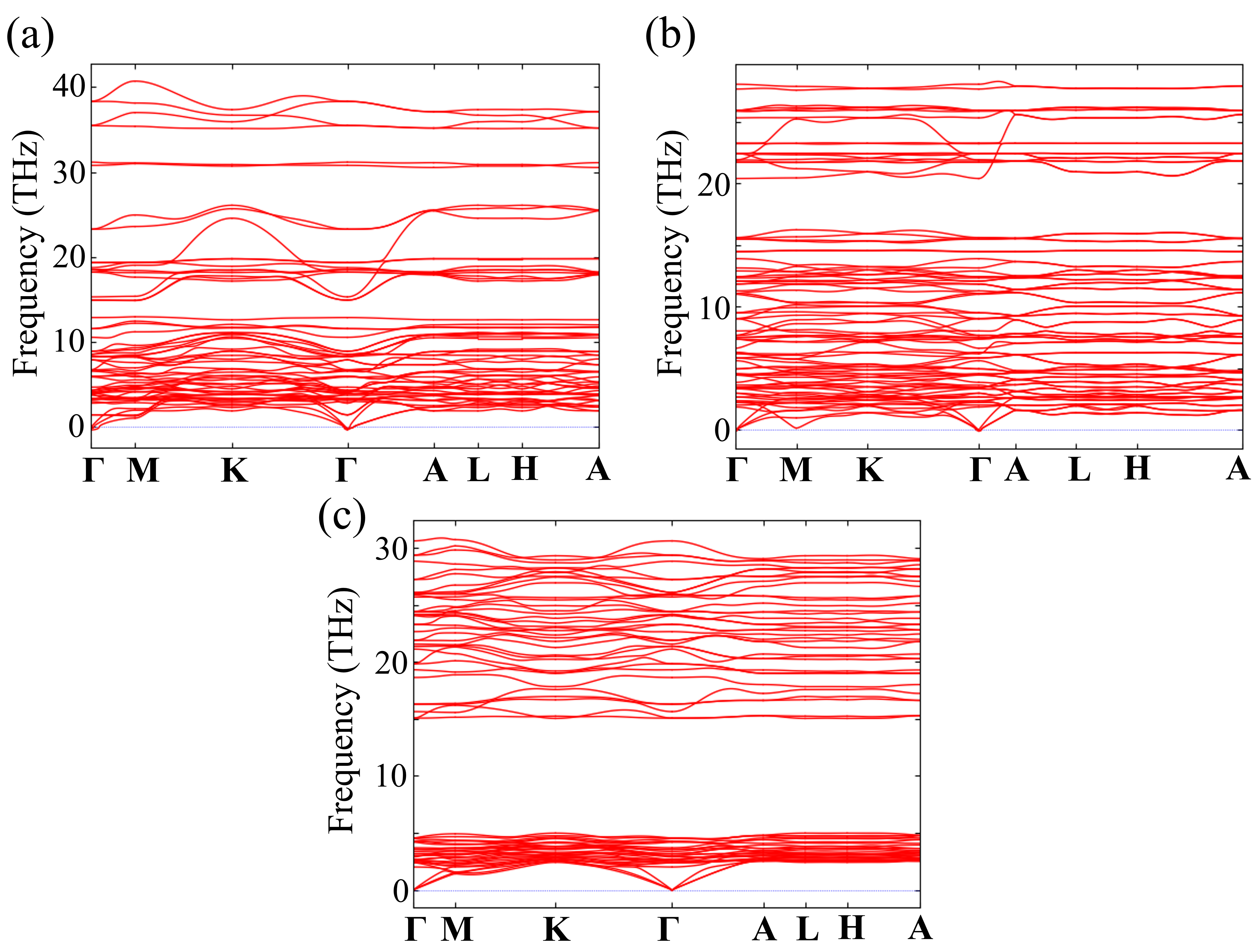}
	\caption{ Phonon band dispersion curves of (a) K$_3$Ta$_3$B$_2$O$_{12}$, (b) KTaGe$_3$O$_{9}$ and (c) Sr$_7$Br$_2$H$_{12}$. }
\end{figure}
\clearpage
\section{E\MakeLowercase{lectronic band structures of} K$_3$T\MakeLowercase{a}$_3$B$_2$O$_{12}$, KT\MakeLowercase{a}G\MakeLowercase{e}$_3$O$_{9}$ \MakeLowercase{and} S\MakeLowercase{r$_7$}B\MakeLowercase{r$_2$}H$_{12}$}
Figure S5 shows the atom projected electronic band structure  for K$_3$Ta$_3$B$_2$O$_{12}$ without including spin orbit coupling (SOC), with SOC using PBE $\epsilon_{xc}$ functional and with SOC using hybrid HSE06 $\epsilon_xc$ functional.  Lower conduction bands (VB) are mainly consists for Ta-5d orbitals while 0-2p orbitals contribute to upper valence bands (VB). Inclusion of SOC leads to splitting of mainly lower CB particularly due to large SOC coming from heavier Ta-5d orbitals. Also, band edge positions are fixed irrespective of choice of functional. Figure S6, S7, S8 shows spin projected band structure for K$_3$Ta$_3$B$_2$O$_{12}$, KTaGe$_3$O$_{9}$ and Sr$_7$Br$_2$H$_{12}$, respectively. For KTaGe$_3$O$_{9}$ and Sr$_7$Br$_2$H$_{12}$, upper valence bands are getting splitted with inclusion of SOC. The band dispersion of splitted bands in $\Gamma-$M$-$K plane is fully governed by the $z$-component of the spin, as discussed in main text in details. 
\begin{figure}[h!]
	\centering
	\includegraphics[width=8.5cm]{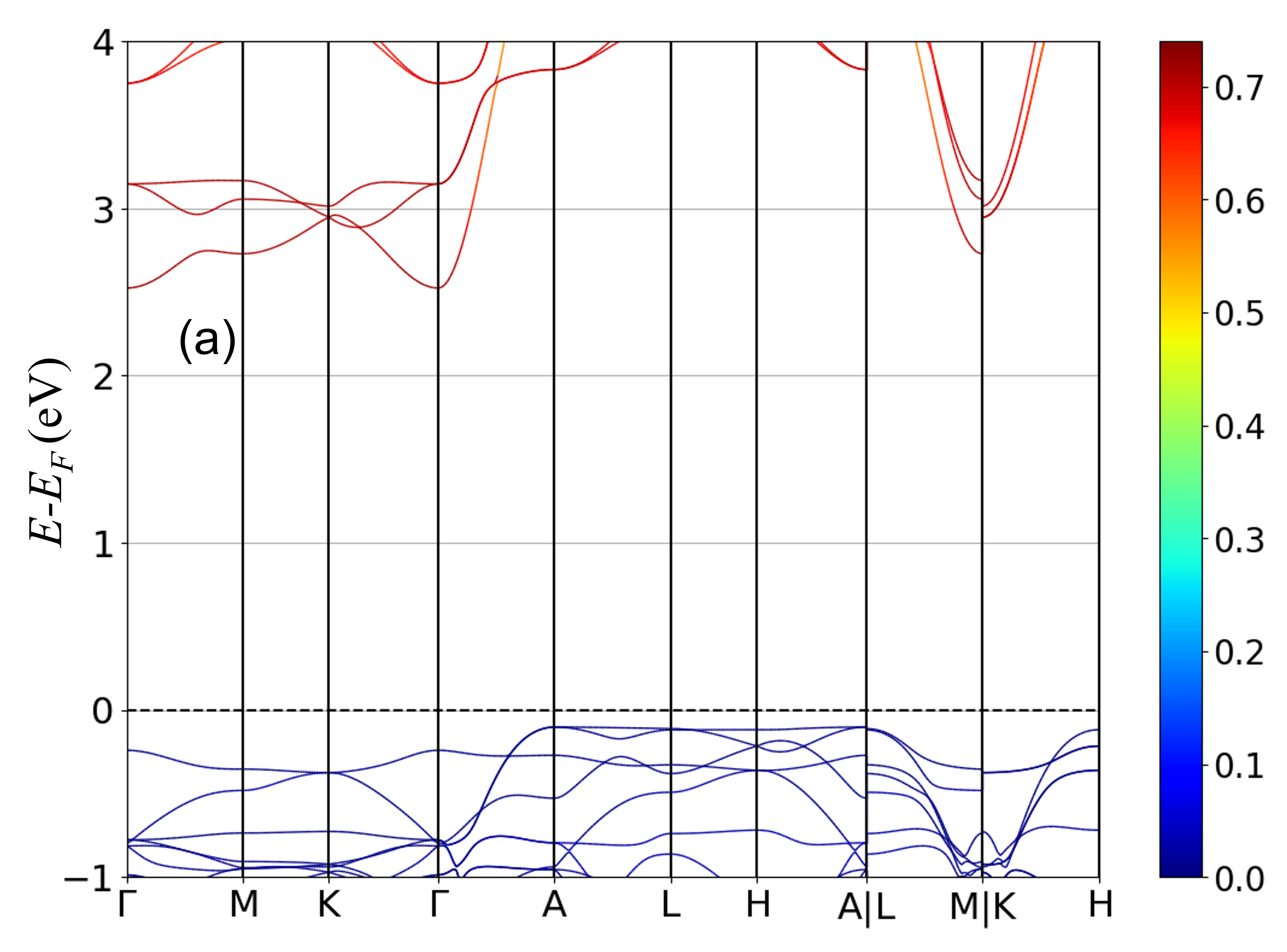}
	\includegraphics[width=8.5cm]{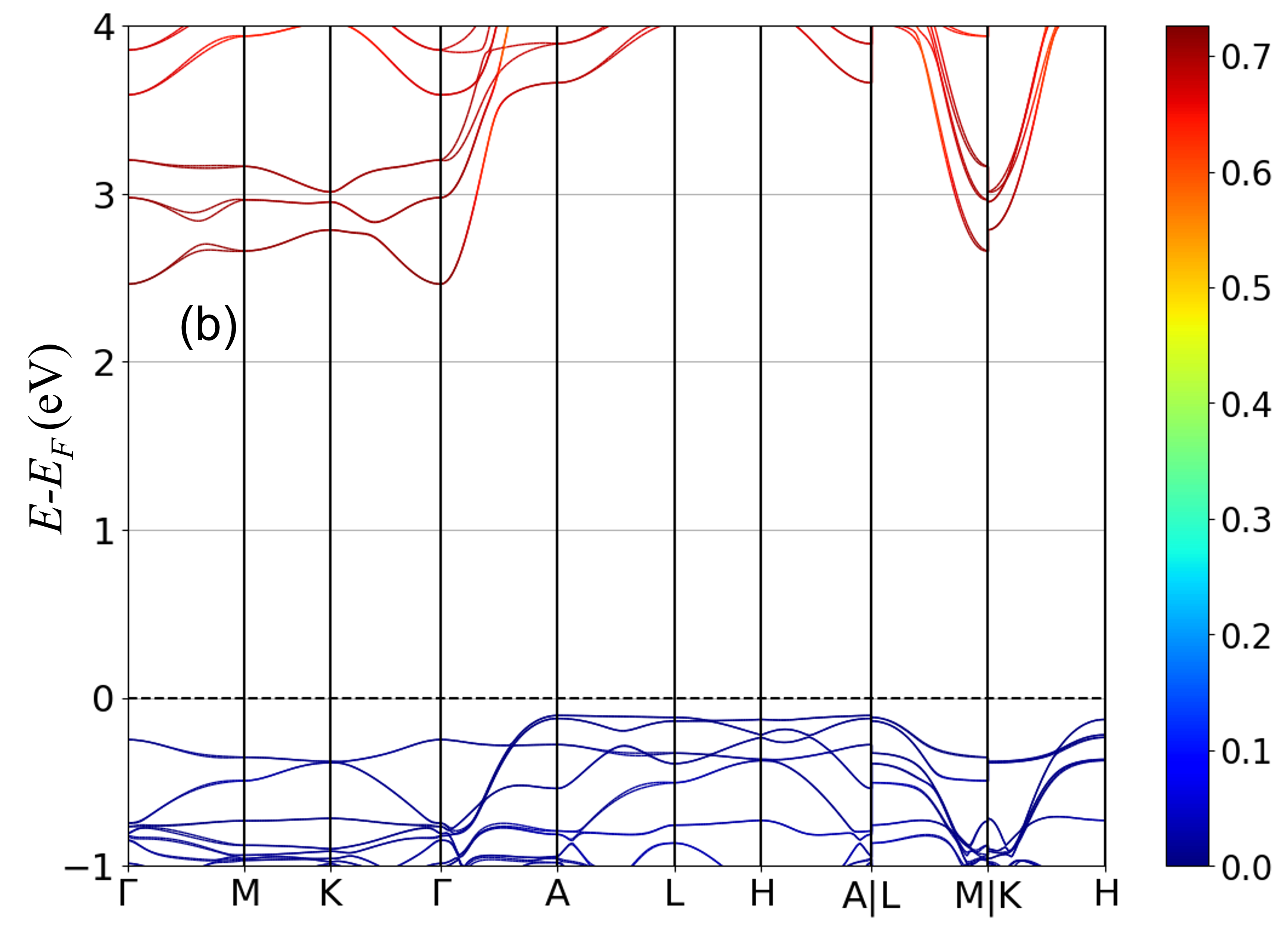}
	\includegraphics[width=8.5cm]{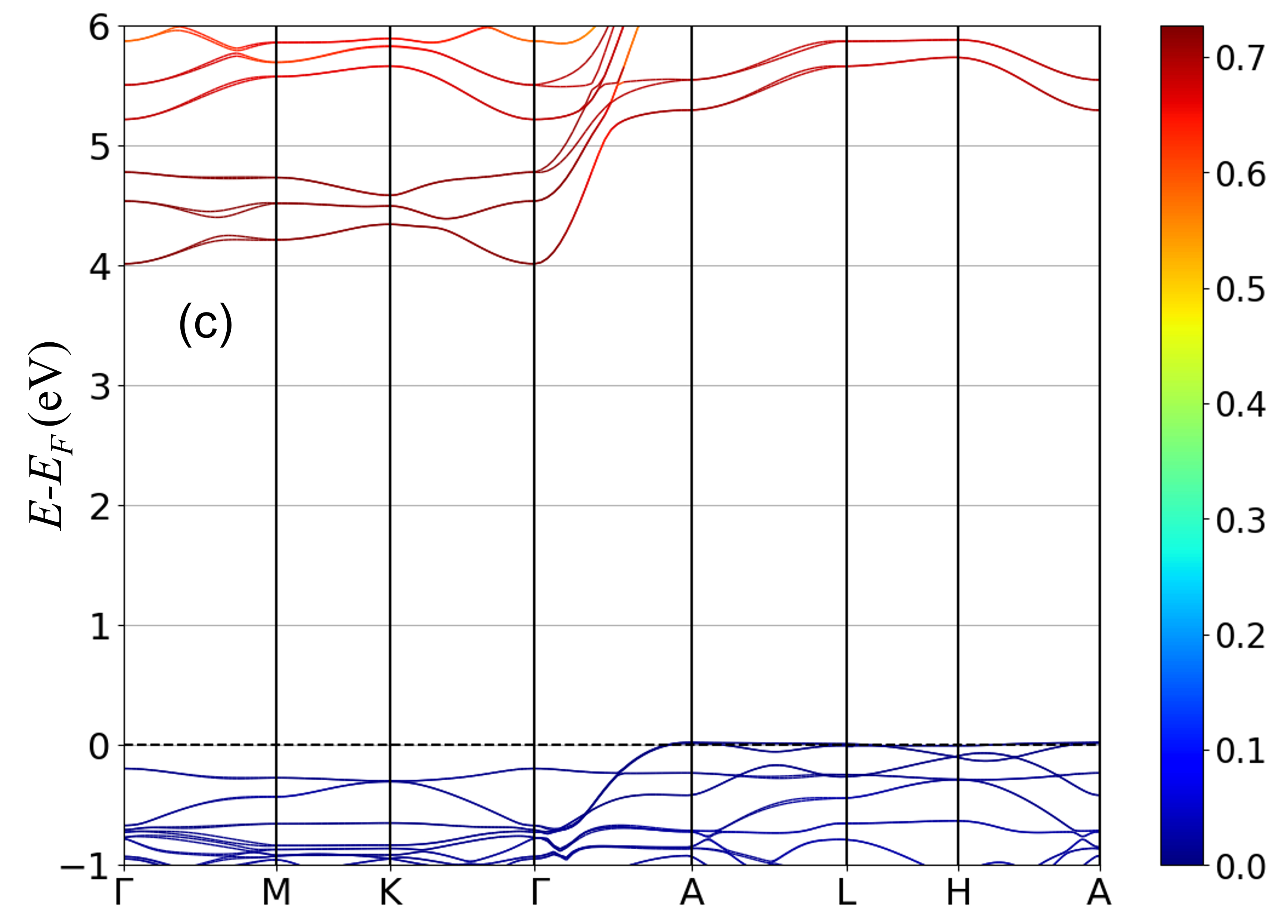}
	\caption{The band structures of K$_3$Ta$_3$B$_2$O$_{12}$ using (a) PBEsol without SOC (b) PBEsol with SOC and (c) HSE06 with SOC projected on Ta-5d orbitals.  }
\end{figure}
\begin{figure}[h!]
	\includegraphics[width=10cm]{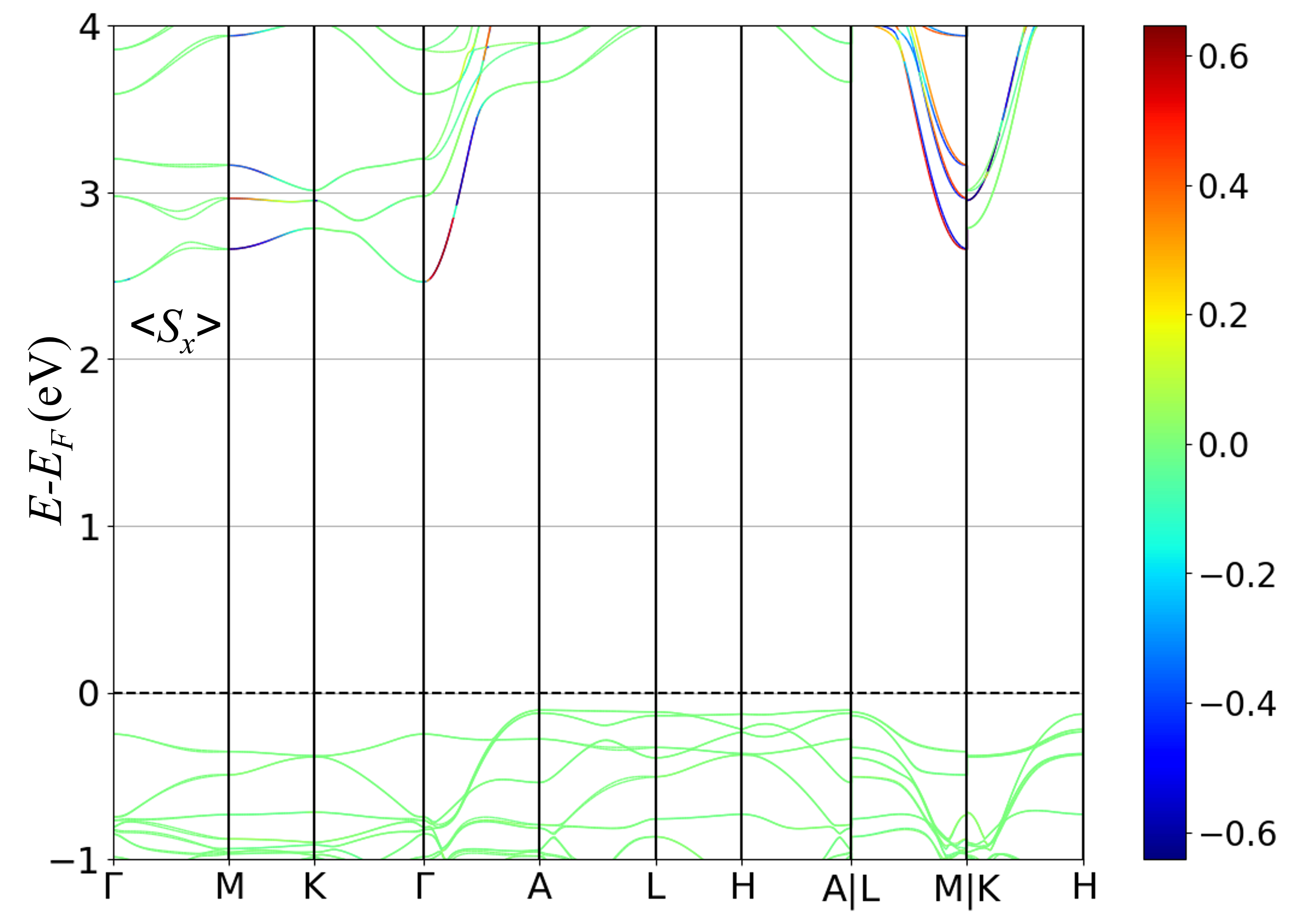}
	\includegraphics[width=10cm]{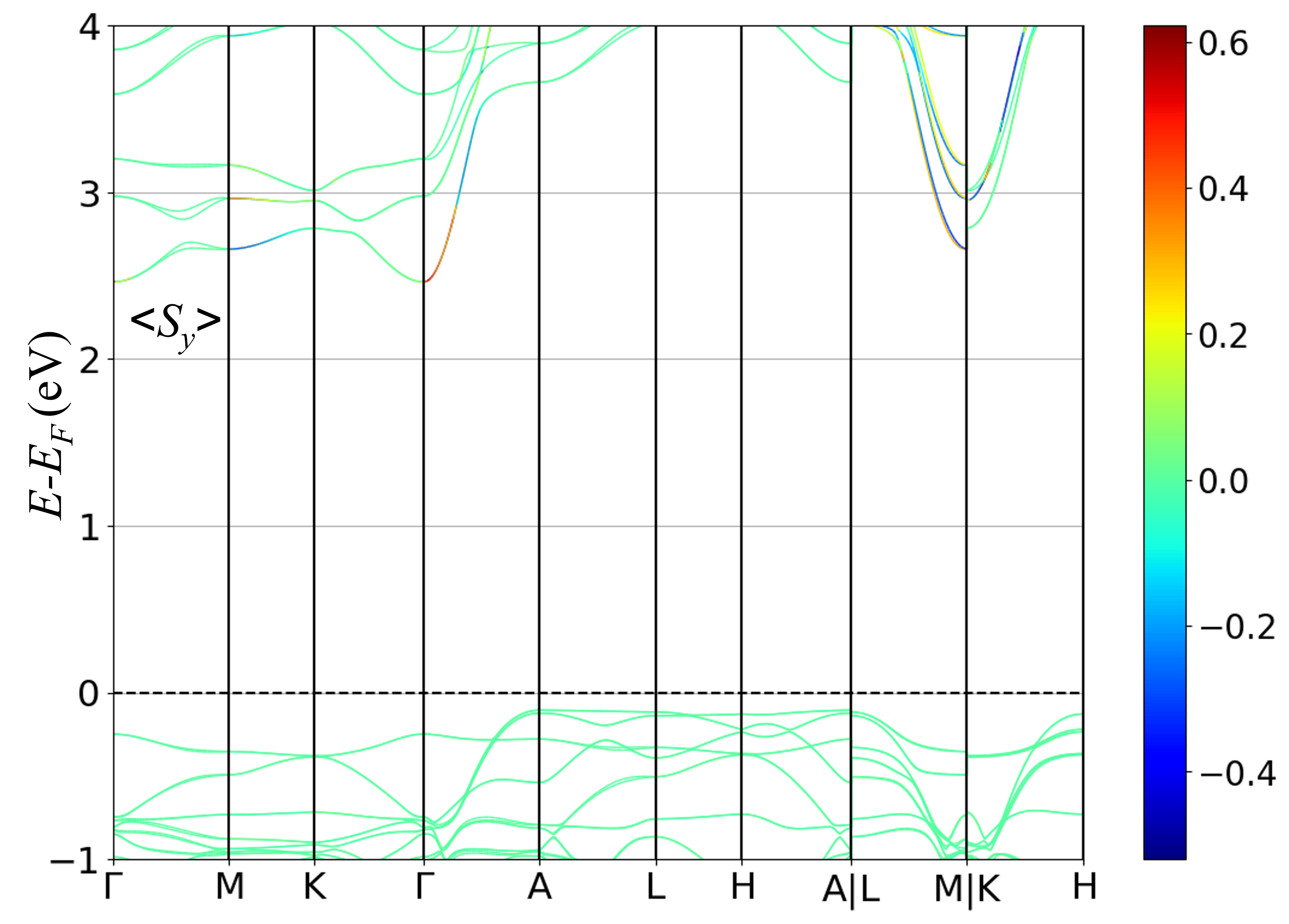}
	\includegraphics[width=10cm]{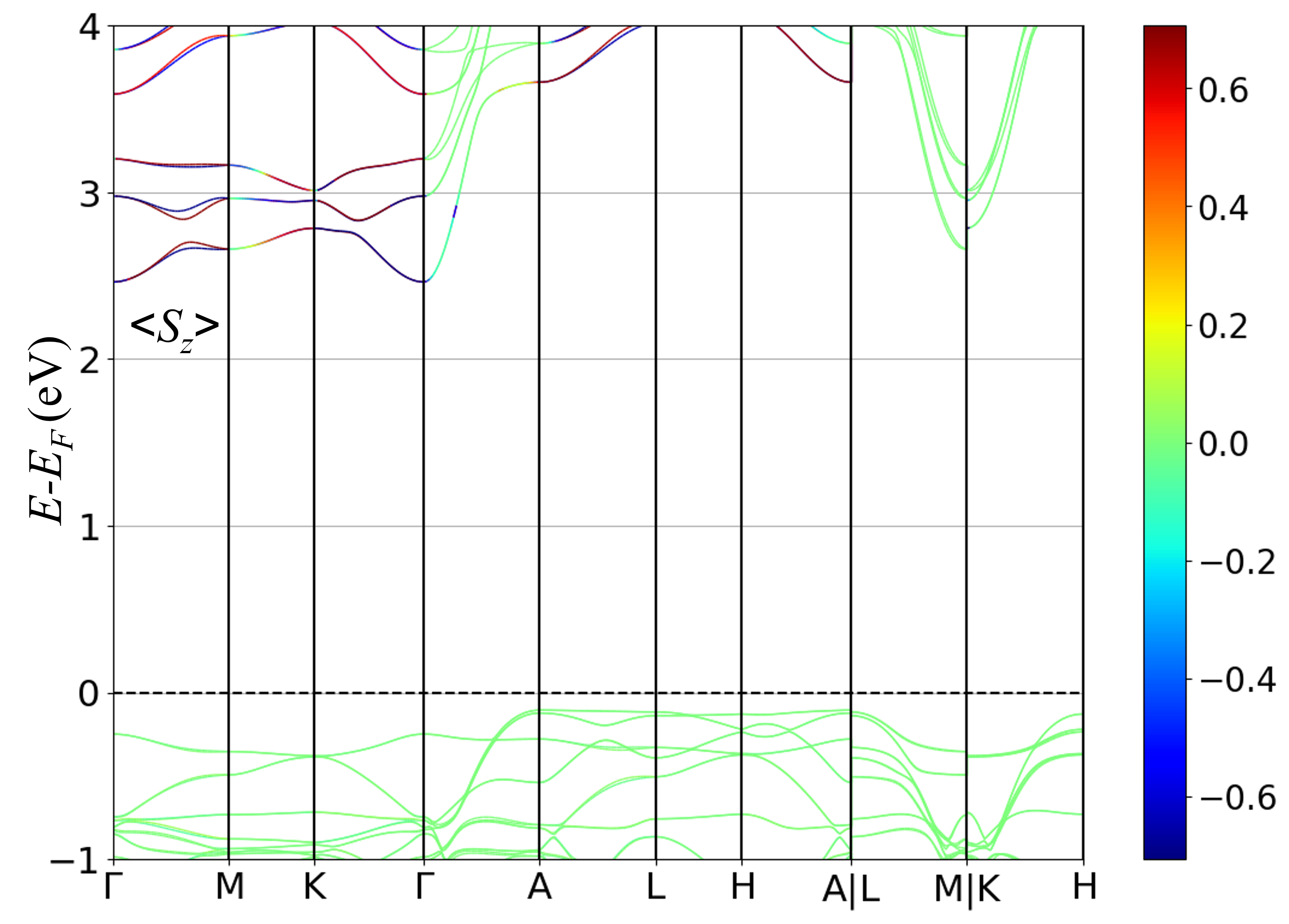}
	\caption{The band structures of K$_3$Ta$_3$B$_2$O$_{12}$ using PBEsol without SOC. Color bars represent expectation values of the spin components $\langle S_x \rangle$, $\langle S_y \rangle$ and $\langle S_z \rangle$. }
\end{figure}
\begin{figure}[h!]
	\includegraphics[width=10cm]{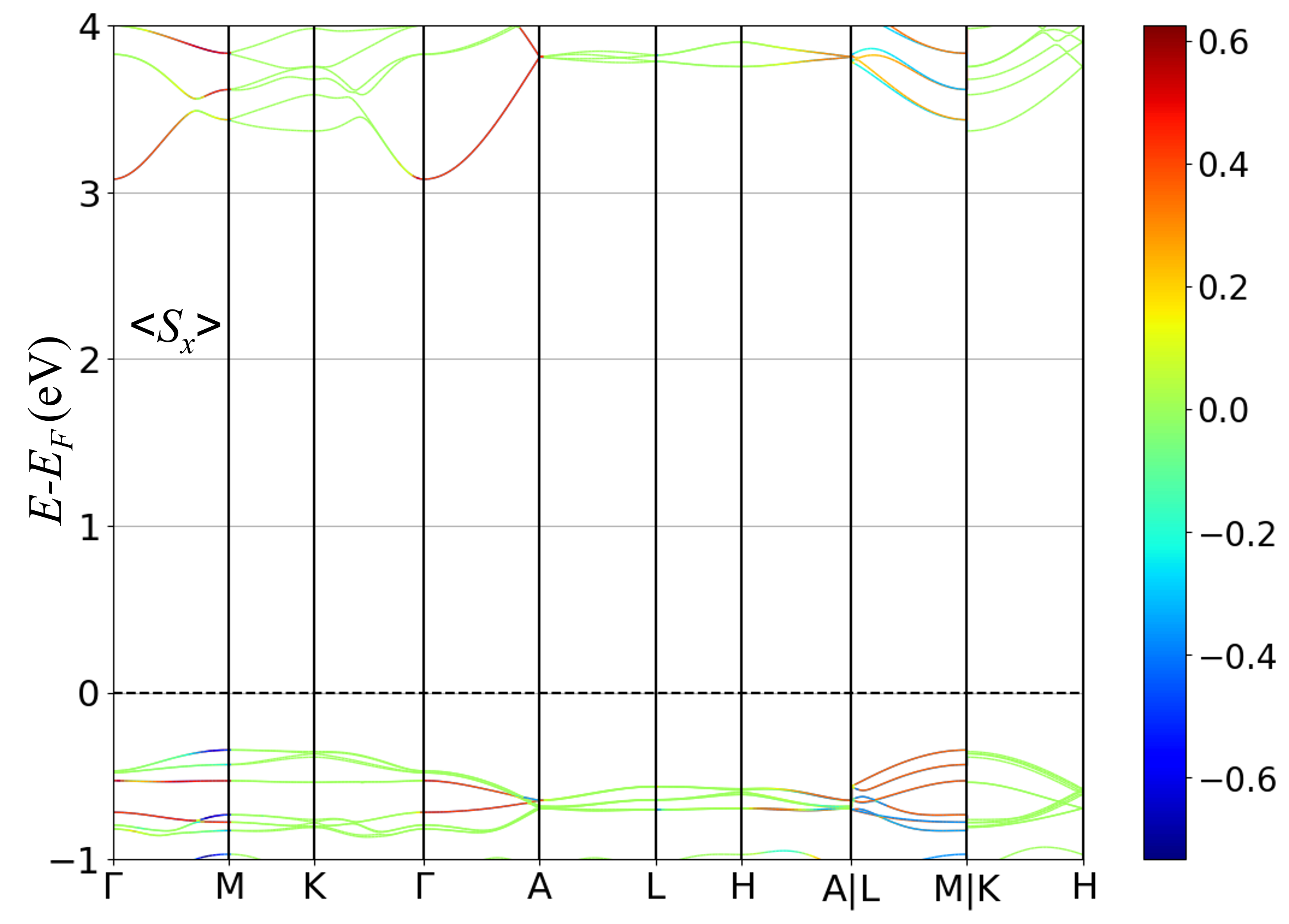}
	\includegraphics[width=10cm]{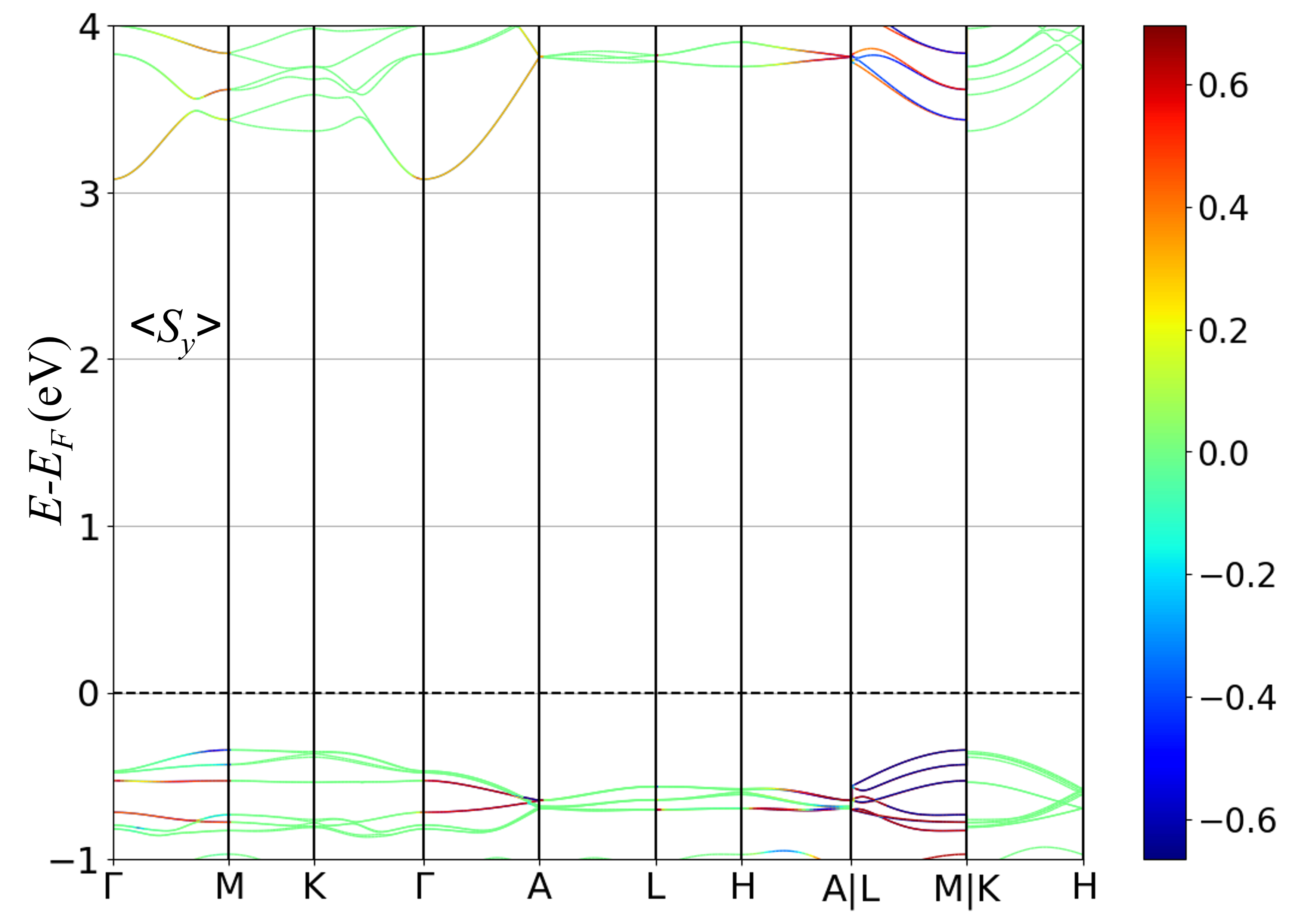}
	\includegraphics[width=10cm]{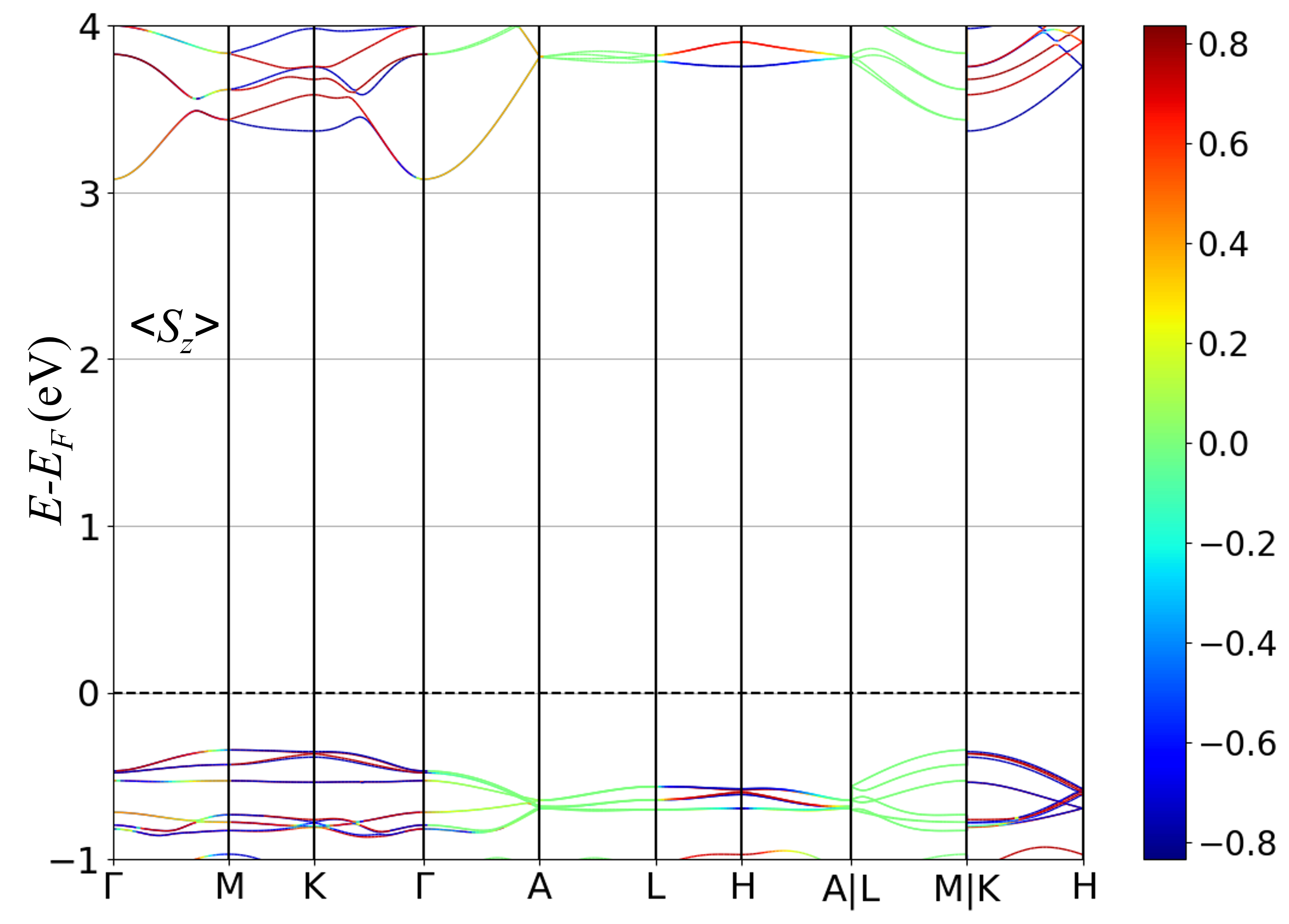}
	\caption{The band structures of KTaGe$_3$O$_{9}$ using PBEsol without SOC. Color bars represent expectation values of the spin components $\langle S_x \rangle$, $\langle S_y \rangle$ and $\langle S_z \rangle$. }
\end{figure}
\begin{figure}[h!]
	\includegraphics[width=10cm]{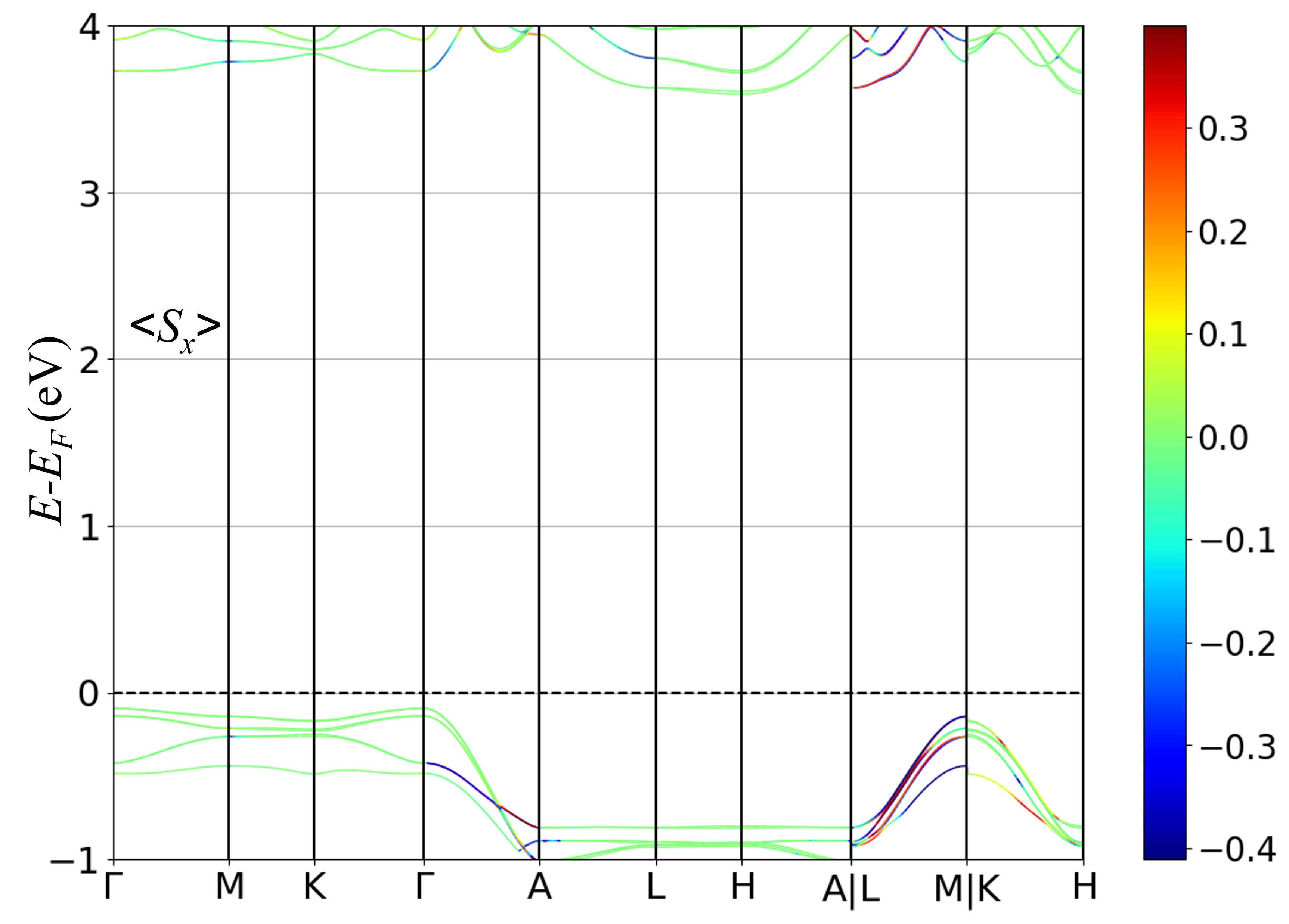}
	\includegraphics[width=10cm]{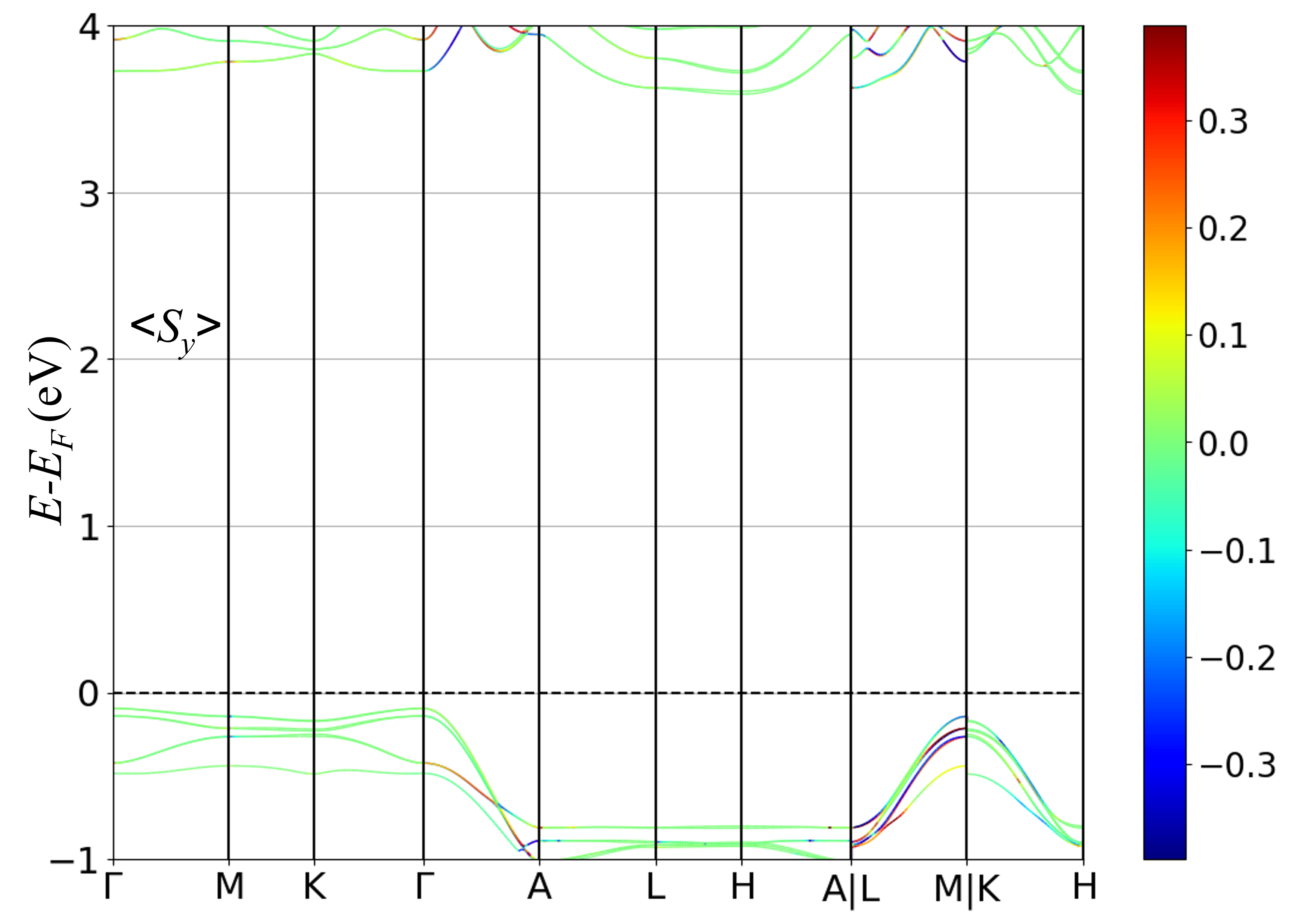}
	\includegraphics[width=10cm]{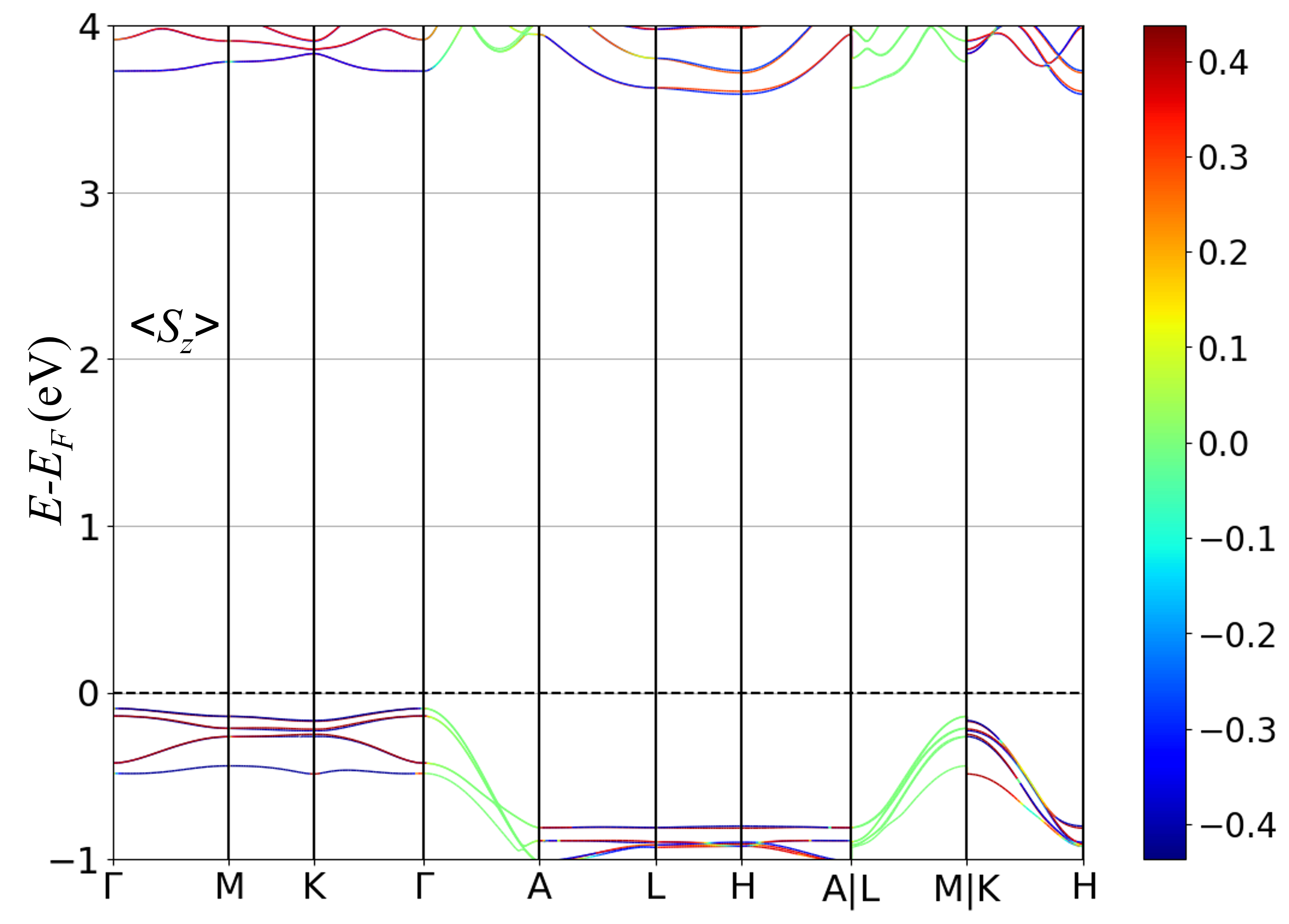}
	\caption{The band structures of Sr$_7$Br$_2$H$_{12}$ using PBEsol without SOC. Color bars represent expectation values of the spin components $\langle S_x \rangle$, $\langle S_y \rangle$ and $\langle S_z \rangle$.}
\end{figure}
\newpage
\section{C\MakeLowercase{omparison between} HSE06, PBE, PBE\MakeLowercase{sol and} PBE+U \MakeLowercase{exchange correlation functionals}} 
In practical, the choice of functional strongly depends upon the chemical system and type of calculations to be performed. Our band structure calculations (see Figure S5) confirm that PBEsol underestimates the band gap as compared to HSE06 functional. Here we have compared the HSE06, PBE, PBEsol and PBE+U with respect to the spin splitting. We have computed the band gap, mass terms ($\alpha$) and cubic splitting parameters ($\zeta$, $\lambda$) with different functionals. Figure S9 and S10 show the DFT resulted spin splittings and spin textures around the $\Gamma$ point of K$_3$Ta$_3$B$_2$O$_{12}$. These spin textures are calculated using 15$\times$15 $\textbf{k}$-mesh around $\Gamma$ point, thus these can be slightly different from the spin textures in main text. HSE06 gives the largest gap of 3.78 eV and PBEsol gives the smallest gap of 2.34 eV (see Table S3). The $\alpha$, $\zeta$ and $\lambda$ obtained using different functionals are nearly same ($\pm 5\%$ with respect to PBEsol) (see Table S3). Since PBEsol is computationally efficient as compared to HSE06, our mostly calculations are based on the PBEsol. 

\begin{figure}[H]
	\begin{center}
	\includegraphics[width=15cm]{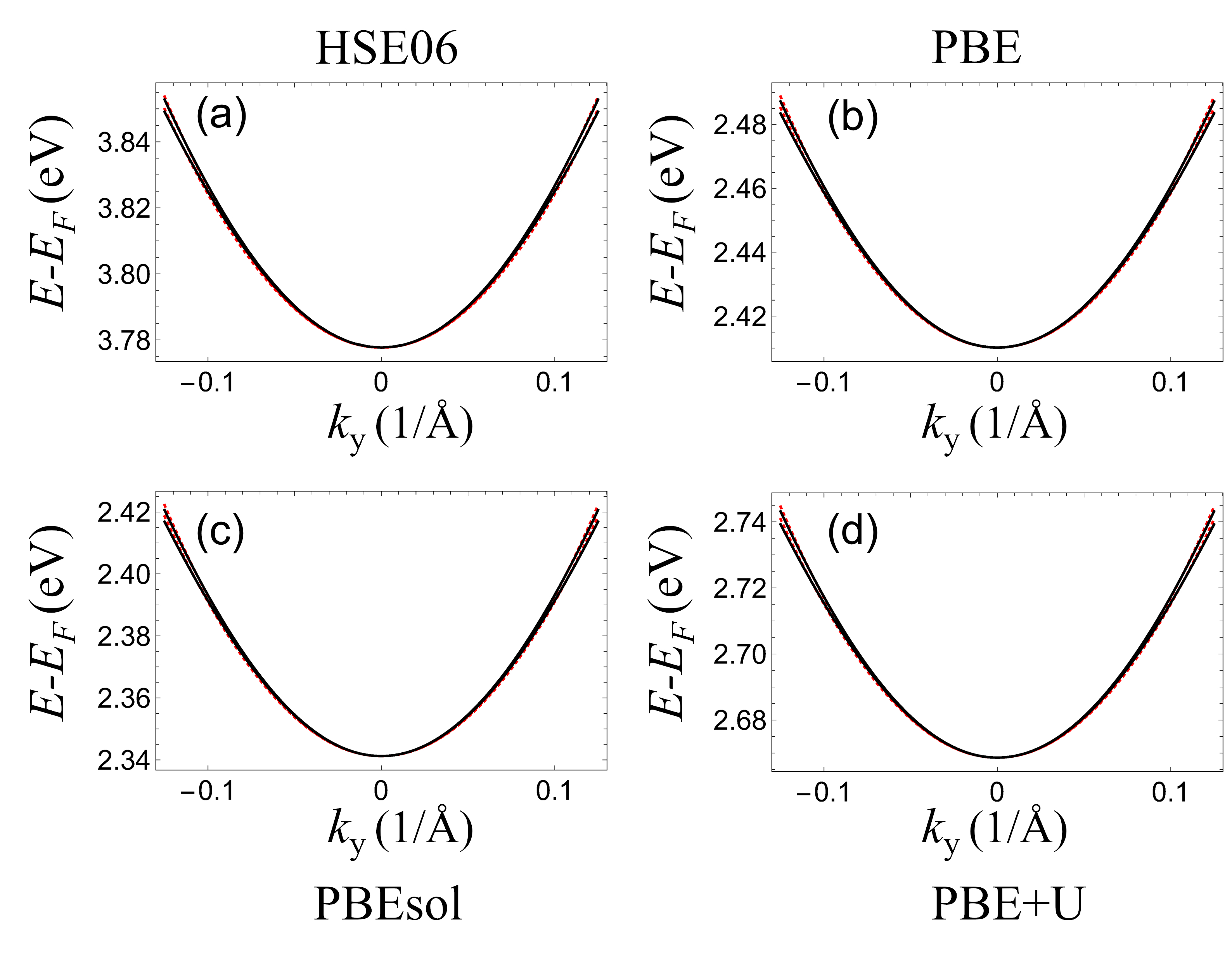}
	\caption{Band structures around the $\Gamma$ point of K$_3$Ta$_3$B$_2$O$_{12}$ along $k_y$ direction calculated using (a) HSE06, (b) PBE, (c) PBEsol and (d) PBE+U exchange correlation functional. The solid black lines and red dots are from the DFT and models, respectively. }
    \end{center}
\end{figure}

\begin{figure}[H]
	\includegraphics[width=18cm]{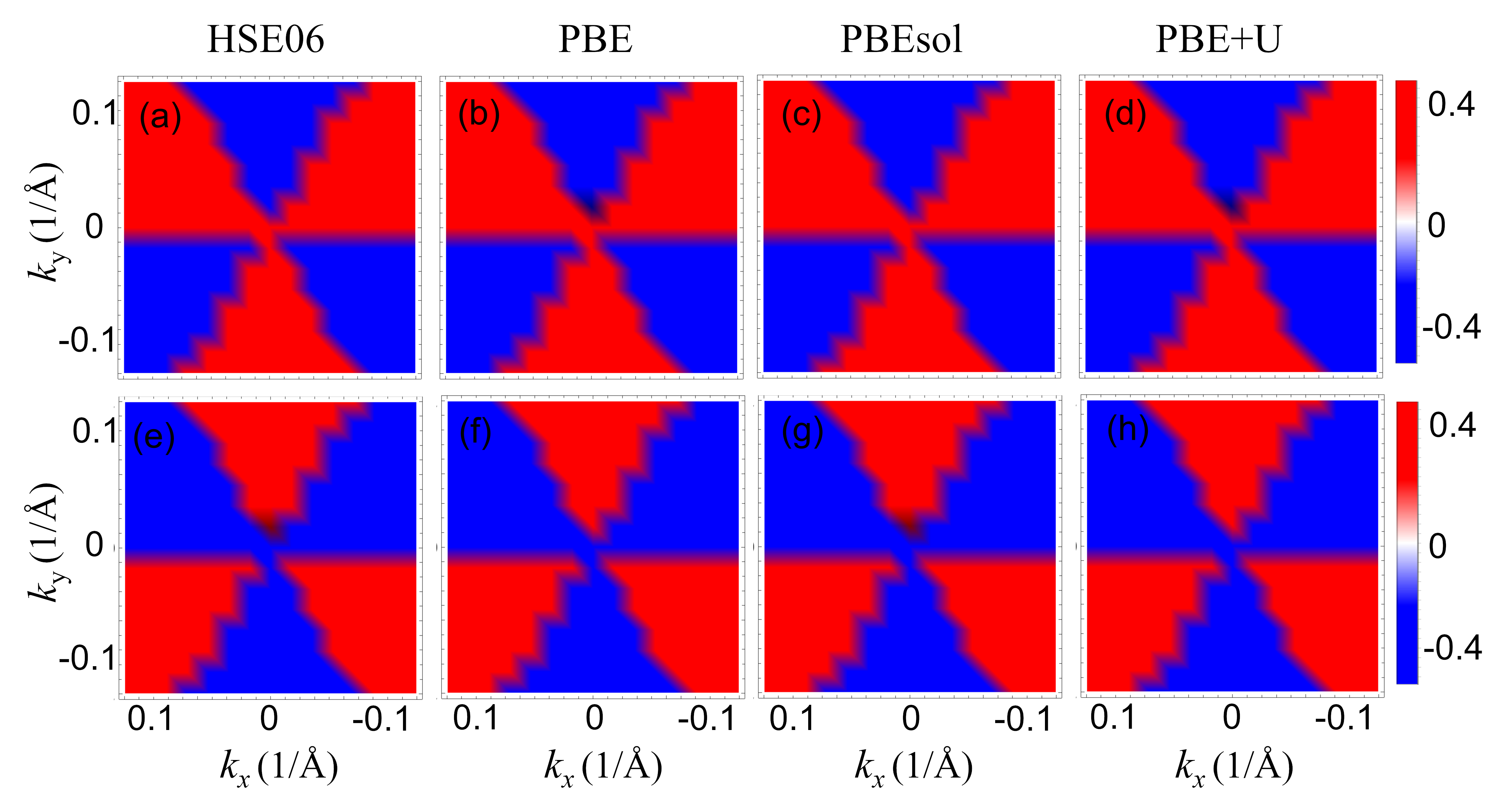}
	\caption{Spin texture of inner conduction bands calculated using DFT with (a) HSE06, (b) PBE, (c) PBEsol and (d) PBE+U exchange correlation functional around the $\Gamma$ point for K$_3$Ta$_3$B$_2$O$_{12}$. Similarly, [(e)-(h)] show the spin texture of outer conduction band.}
\end{figure}

\begin{table}[H]
	\begin{center}
		\caption{Parametrization of $\alpha$, $\lambda$, and $\eta$ of model for low conduction bands (around the $\Gamma$ point) of K$_3$Ta$_3$B$_2$O$_{12}$. The $k_x$ and $k_y$ ranges for fitting our models are 0.125 \AA$^{-1}$.} 
		\label{tbl1}
		\begin{tabular}{ p{4.5 cm} p{4.5 cm} p{4.5 cm} p{2 cm}  }
			\hline
			\hline
			{Functional}     &  {Band gap (eV)}   & {$\alpha$ (eV\AA$^2$)} &   {$\lambda$ (eV\AA$^3$)}  \\ \hline
			HSE06  & 3.78  &  4.78              & 6.70  \\  
			PBE    & 2.41  &  4.91  &  6.78  \\
			PBEsol & 2.34  &  5.07  &   6.85 \\   
			PBE+U  & 2.67  &  4.74  &  7.14  \\ 
			\hline\hline
			
		\end{tabular}
	\end{center}
\end{table}
\newpage
\section{S\MakeLowercase{pin textures}}
Figure S11 shows the persistent spin textures computed using DFT and $k.p$ model Hamiltonian computed using the same dense $k-$grid. The boarder lines separating the $+|s_z|$ and $-|s_z|$ vary case by case and are estimated using model as shown in main text. The DFT and model resulted band structures are in well agreement with each other.
\begin{figure}[h!]
	    \includegraphics[width=18cm]{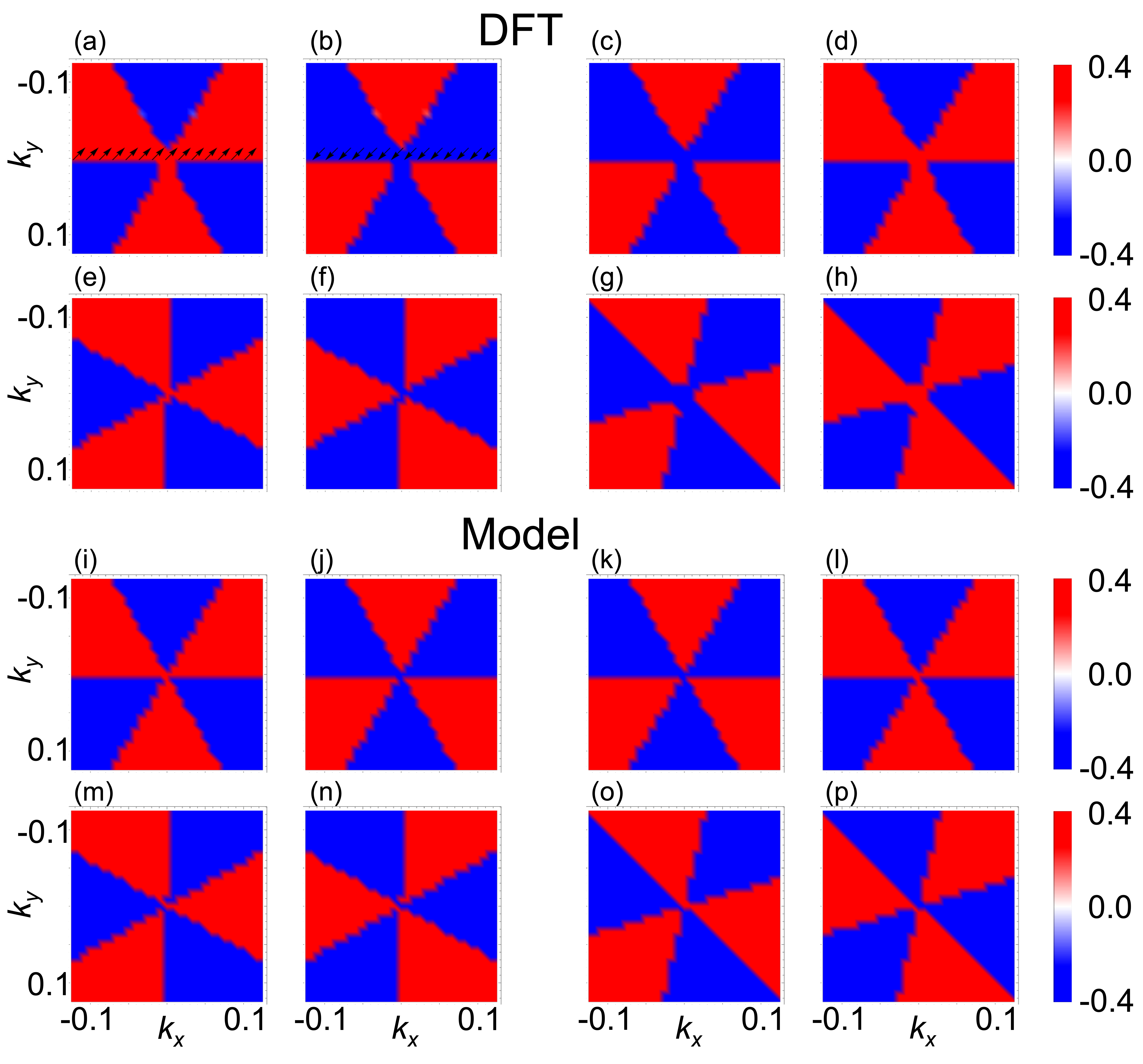}
		\caption{Spin textures of the conduction bands for K$_3$Ta$_3$B$_2$O$_{12}$ around the $\Gamma$ [(a)-(b)] and A[(c)-(d)] points obtained by DFT. Spin textures of the valence bands for KTaGe$_3$O$_{9}$ [(e)-(f)] and Sr$_7$Br$_2$H$_{12}$ [(g)-(h)] around the $\Gamma$ point obtained by DFT. (i)-(p) Counterparts of (a)-(h) obtained using $k.p$ model Hamiltonian. The arrows and color bars denote the in-plane and out-of-plane components of spin textures with respect to the $(k_x-k_y)$ plane. The units of $k_x$ and $k_y$ are \AA$^{-1}$.}
		\label{fig3}
	
\end{figure}
\newpage
\section{S\MakeLowercase{tability under strain}}
\begin{figure}[h!]
	\includegraphics[width=15cm]{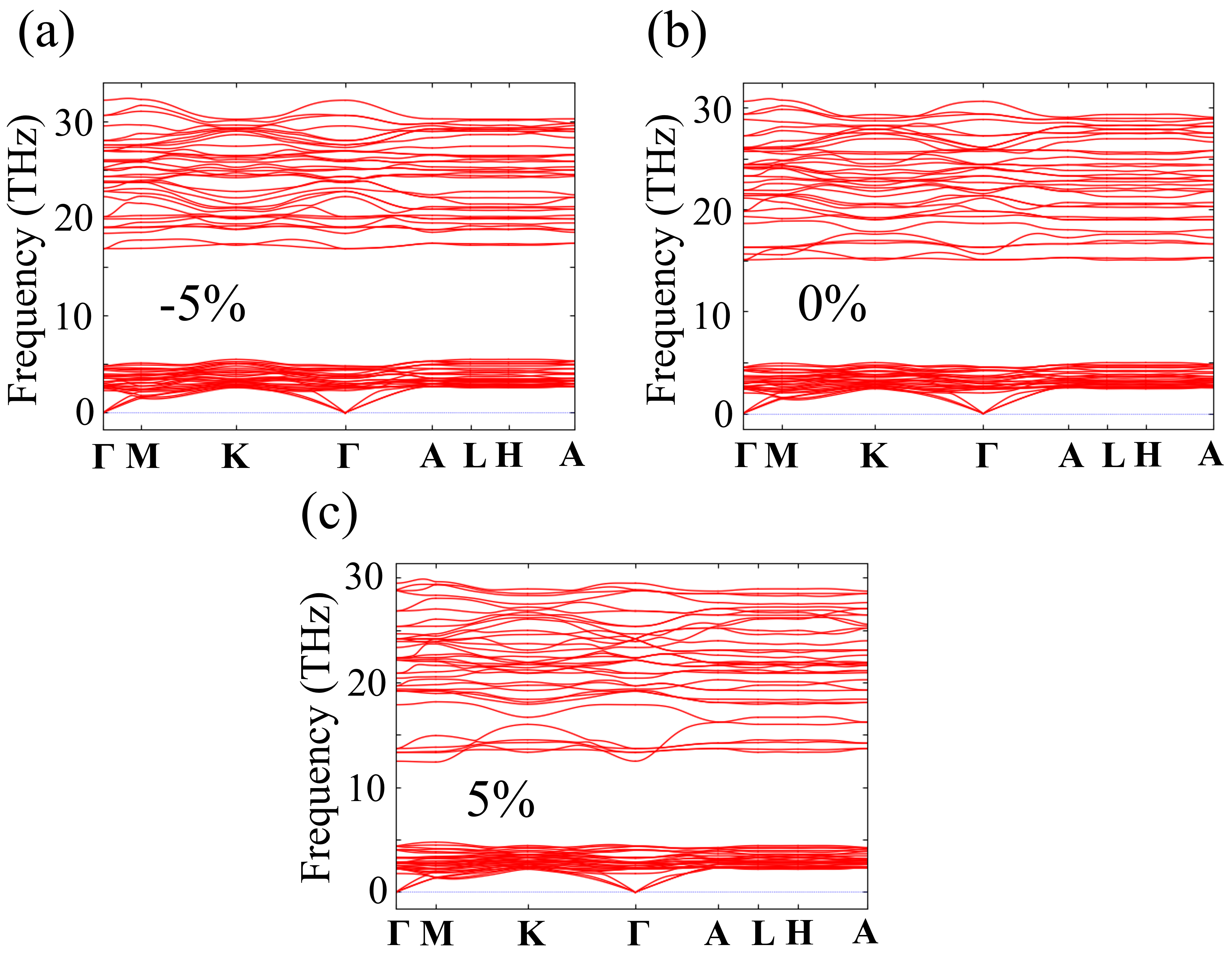}
	\caption{Phonon band dispersion curves of Sr$_7$Br$_2$H$_{12}$ under (a) -5\%, (b) 0\% and (c) 5\% uniaxial strain.}
\end{figure}

\clearpage
\section*{References}
\bibliography{sci}